\documentclass[hyper]{JHEP3} 

\usepackage{axodraw}
\usepackage{epsfig}
\usepackage{amsmath}
\usepackage{cite}

\newcommand{\beq}{\begin{equation}}
\newcommand{\eeq}{\end{equation}}
\newcommand{\emiss}{\rlap{\,/}E}

\title{Contrasting Supersymmetry and Universal Extra Dimensions
           at the CLIC Multi-TeV $e^+e^-$ Collider}

\author{Marco Battaglia\\
        Dept. of Physics, University of California, Berkeley, CA 94720, USA\\
	and\\
	Lawrence Berkeley National Laboratory, Berkeley, CA 94720, USA\\
        E-mail: \email{MBattaglia@lbl.gov} }
\author{AseshKrishna Datta\\
        MCTP, University of Michigan, 
        Ann Arbor, MI 48109-1120, USA\\
        E-mail: \email{asesh@umich.edu} }
\author{Albert De Roeck\\
        CERN, Geneva, Switzerland\\
        E-mail: \email{Albert.de.Roeck@cern.ch} }
\author{Kyoungchul Kong\\
        Physics Dept., University of Florida,
        Gainesville, FL 32611, USA\\
        E-mail: \email{kong@phys.ufl.edu} }
\author{Konstantin T.~Matchev\\
        Physics Dept., University of Florida,
        Gainesville, FL 32611, USA\\
        E-mail: \email{matchev@phys.ufl.edu}}

\received{\today}               
\accepted{\today}               

\preprint{UFIFT-HEP-05-05 \\
          MCTP-05-42 
          } 

\abstract{Universal extra dimensions and supersymmetry have rather similar
experimental signatures at hadron colliders. 
The proper interpretation of an LHC discovery in either case may 
therefore require further data from a lepton collider. In this paper we identify 
methods for discriminating between the two scenarios at the linear collider. 
We study the processes of Kaluza-Klein muon pair production in universal extra
dimensions in parallel to smuon pair production in supersymmetry, accounting for the 
effects of detector resolution, beam-beam interactions and accelerator induced 
backgrounds. We find that the angular distributions of the final state muons,
the energy spectrum of the radiative return photon and the 
total cross-section measurement are powerful discriminators
between the two models. Accurate determination of the particle masses can be 
obtained both by a study of the momentum spectrum of the final state leptons and 
by a scan of the particle pair production thresholds. We also calculate the
production rates of various Kaluza-Klein particles and discuss the
associated signatures.}


\begin{document} 

\section{Introduction}

Supersymmetry (SUSY) and Extra Dimensions offer two different paths to a theory of new 
physics beyond the Standard Model (SM). They both address the hierarchy problem, 
play a role in a more fundamental theory aimed at unifying the SM with gravity, and 
offer a candidate particle for dark matter, compatible with present cosmology data.
If either supersymmetry or extra dimensions exist at the TeV scale, signals of new 
physics should be found by the ATLAS and CMS experiments at the Large Hadron 
Collider (LHC) at CERN. However, as we discuss below, 
the proper interpretation of such discoveries, 
namely the correct identification of the nature of the new physics signals, may not
be straightforward at the LHC and may require the complementary data from an 
$e^+e^-$ collider. In particular, a multi-TeV collider, such as 
CLIC~\cite{Assmann:2000hg,Group:2004sz}, would ensure a sensitivity over a broad mass range.

A particularly interesting scenario of TeV-size extra dimensions is 
offered by the so called Universal Extra Dimensions (UED),
originally proposed in~\cite{Appelquist:2000nn}, where all SM
particles are allowed to freely propagate into the bulk. 
The case of UED bears interesting analogies
to supersymmetry\footnote{More precisely, the phenomenology of
the first level ($n=1$) of Kaluza-Klein modes in UED is very similar 
to that of $N=1$ supersymmetric models with somewhat degenerate superpartner spectrum
and stable lightest supersymmetric particle (LSP). 
In what follows we shall use the term ``supersymmetry''
in this somewhat narrower context.}, and sometimes has been
referred to as  ``bosonic supersymmetry'' \cite{Cheng:2002ab}. In principle, 
disentangling UED and supersymmetry appears highly non-trivial at hadron 
colliders~\cite{Cheng:2002ab,Rizzo:2001sd,Macesanu:2002db}.
For each SM particle, both models predict the existence of
a partner (or partners) with identical interactions.
Unfortunately, the masses of these new particles are model-dependent
and cannot be used to unambiguously discriminate between the two theories,
although a degenerate spectrum might suggest UED while 
a split spectrum would hint\footnote{Notice that 
the recently proposed little Higgs models with $T$-parity 
\cite{Cheng:2003ju,Cheng:2004yc,Hubisz:2004ft} are reminiscent of UED,
and their spectrum does not have to be degenerate, so they
may still be confused with supersymmetry. Under those circumstances, the
methods for discrimination discussed in this paper would still apply.} 
towards SUSY. One would therefore like 
to have an experimental verification which relies on the fundamental distinctions 
between the two models.

One of the characteristic features of UED is the presence
of a whole tower of Kaluza-Klein (KK) partners, labelled by their KK level $n$.
In contrast, $N=1$ supersymmetry predicts a single superpartner for each SM particle.
One might therefore hope to be able to discover the higher KK modes of UED
and, having observed a repetition of the SM particle content at each KK level,
prove the existence of extra dimensions. However, there are two significant 
challenges along this route. First, the masses of the higher KK modes are (roughly)
integer multiples of the masses of the $n=1$ KK partners, and as a result 
their production cross-sections are kinematically suppressed. Second, they 
predominantly decay to $n=1$ KK modes, thus simply contributing a small amount to
the inclusive production of $n=1$ KK particles. Just as in the case of SUSY, 
because of the unknown momentum carried away by the dark matter candidate
at the end of the decay chain, one is unable to reconstruct individual
KK resonances. The only exceptions are the level $2$ KK gauge bosons, 
which may appear as high mass dijet or dilepton resonances, when they 
decay directly to SM fermions through loop suppressed
couplings \cite{Cheng:2002ab,Georgi:2000ks,vonGersdorff:2002as,Cheng:2002iz}.
Unfortunately the reach of the LHC is rather limited in this case~\cite{DKM},
due to the smallness of these loop-suppressed branching fractions.
In addition, even if they are discovered, such 
resonances can be misinterpreted as ordinary $Z'$ gauge bosons in 
extended supersymmetric models~\cite{DKM}.

The second fundamental distinction between SUSY and {\em the minimal version of} 
UED is that supersymmetric extensions of the SM necessarily have 
extended Higgs sectors leading to additional Higgs and higgsino states
in the spectrum. The higgsinos of supersymmetry are in one-to-one 
correspondence with the $n=1$ KK partners of the Goldstone bosons
and cannot be used for discrimination. The additional Higgs bosons
of SUSY are not a robust discriminator either.
On the one hand, they may simply escape detection at the LHC:
the SUSY parameter space has large portions where the LHC discovers 
a single (SM-like) Higgs boson and misses the rest of the Higgs spectrum. 
Alternatively,
one may consider UED models with an extended Higgs sector
which would mimic the SUSY Higgs phenomenology.

The third and most fundamental distinction between UED and supersymmetry is 
reflected in the properties of the individual particles: the
KK partners have identical spin quantum numbers as their SM counterparts, 
while the spins of the superpartners differ by $1/2$ unit. 
However, spin determinations also appear to be difficult at the LHC
(or at hadron colliders in general), where the center of mass energy
in each event is unknown. In addition, the momenta of the two dark
matter candidates in the event are also unknown. 
This prevents the reconstruction of any
rest frame angular decay distributions, or the directions of the
two particles at the top of the decay chains. Recently it has been
suggested that a charge asymmetry in the lepton-jet invariant mass
distributions from a particular cascade, can be used to discriminate
SUSY from the case of pure phase space decays~\cite{Barr:2004ze}.
However, the possibility of discriminating SUSY and UED by this
method still needs to be demonstrated.

At the LHC, strong processes dominate the production of both KK 
modes and superpartners. The resulting
signatures involve relatively soft jets, leptons and missing transverse
energy~\cite{Cheng:2002ab}. The study in~\cite{Cheng:2002ab} shows that,
with $100\ {\rm fb}^{-1}$ of integrated luminosity, the LHC experiments will be able 
to cover all of the cosmologically preferred parameter space in 
UED~\cite{Servant:2002aq}. However, due to the reasons explained above, 
the discrimination between UED and supersymmetry may be difficult at the LHC.

In this paper, we focus on the minimal UED model, described in some detail in 
Section~\ref{sec:ued}, 
and on the Minimal Supersymmetric extension of the Standard 
Model (MSSM) to study how the two models can be discriminated at a linear collider
tunable over a center-of-mass energy range of 1~TeV to 3~TeV. 
We assume that the LHC will have already observed 
signals of new physics consistent with either $n=1$ KK modes in UED or sparticle 
production in supersymmetry. We concentrate on the relevant questions to be
addressed by a post-LHC facility. In section~\ref{sec:evsim} we describe the 
event simulation and reconstruction adopted in this analysis, while in 
section~\ref{sec:contrast} we contrast UED with supersymmetry using the 
example of $\mu^+\mu^-\emiss$ final state. In section \ref{sec:other}
we extend our discussion to other possible final states.
Section~\ref{sec:conclusions} has our conclusions.

\section{The minimal UED model} 
\label{sec:ued}

In its simplest incarnation, the UED model has all the SM particles
propagating in a single extra dimension of size $R$,
which is compactified on an $S_1/Z_2$ orbifold.
More complicated versions have also been proposed,
motivated by ideas about electroweak symmetry breaking~\cite{Arkani-Hamed:2000hv},
neutrino masses~\cite{Appelquist:2002ft,Mohapatra:2002ug}, 
proton stability~\cite{Appelquist:2001mj} or the number of 
generations~\cite{Dobrescu:2001ae}. A peculiar feature of UED is the conservation of
Kaluza-Klein number at tree level, which is a simple consequence
of momentum conservation along the extra dimension.
However, bulk and brane radiative effects 
\cite{Georgi:2000ks,vonGersdorff:2002as,Cheng:2002iz}
break KK number down to a discrete conserved quantity,
the so called KK parity, $(-1)^n$, where $n$ is the KK level.
KK parity ensures that the lightest KK partners --
those at level one -- are always pair-produced
in collider experiments, similar to the case of supersymmetry models
with conserved $R$-parity. KK parity conservation also implies
that the contributions to various precisely measured low-energy
observables \cite{Agashe:2001ra,Agashe:2001xt,Appelquist:2001jz,%
Petriello:2002uu,Appelquist:2002wb,Chakraverty:2002qk,Buras:2002ej,Buras:2003mk}
only arise at loop level and are small.
As a result, the limits on the scale of the extra dimension, from 
precision electro-weak data, are rather weak, constraining  $R^{-1}$ to 
be larger than approximately 250~GeV \cite{Appelquist:2002wb}. An attractive feature of
UED is the presence of a stable massive particle which can be
a cold dark matter candidate \cite{Dienes:1998vg,Cheng:2002iz,Servant:2002aq,Cheng:2002ej}.
The lightest KK partner (LKP) at level one 
has negative KK parity and can be stable on cosmological scales.
In the minimal model, the LKP is the KK partner of the
hypercharge gauge boson \cite{Cheng:2002iz} and
its relic density is generically in the desired range
\cite{Servant:2002aq}. Kaluza-Klein dark matter offers 
excellent prospects for direct or indirect detection
\cite{Cheng:2002ej,Hooper:2002gs,Servant:2002hb,Majumdar:2002mw,Bertone:2002ms,Hooper:2004xn,Bergstrom:2004cy}.
Once the radiative corrections to the Kaluza-Klein masses are 
properly taken into account \cite{Cheng:2002iz},
the collider phenomenology of the minimal UED model\footnote{For
alternative possibilities, see \cite{Macesanu:2002ew}.}
exhibits striking similarities to supersymmetry
\cite{Cheng:2002ab,Cheng:2002rn}
and represents an interesting and well motivated
counterexample which can ``fake'' supersymmetry signals at the LHC.

In the minimal UED model, the bulk interactions of the KK modes readily follow 
from the SM Lagrangian and contain no unknown parameters other
than the mass, $M_H$, of the SM Higgs boson. In contrast, the boundary 
interactions, which are localized on the orbifold fixed points, are in 
principle arbitrary, and thus correspond to new free parameters in the theory. 
They are in fact renormalised
by bulk interactions, and are scale dependent \cite{Georgi:2000ks}.
Therefore, we need an ansatz for their values at a particular scale.
Of course, the UED model should be treated only as an effective
theory which is valid up to some high scale $\Lambda$, at which
it is matched to some more fundamental theory. In the minimal
UED model the boundary terms are assumed to vanish at the 
cutoff scale $\Lambda$, and are subsequently generated through
RGE evolution to lower scales. Thus the minimal UED model 
has only two input parameters: the size of the extra dimension, $R$, 
and the cutoff scale, $\Lambda$. The number of
KK levels present in the effective theory is therefore $\Lambda R$
and may vary between a few and $\sim 40$, where the upper limit corresponds to 
values of $\Lambda$ leading to a breakdown of perturbativity below the $\Lambda$ scale.

\section{Event simulation and data analysis}
\label{sec:evsim}

In order to study the discrimination of UED signals from supersymmetry,
we have implemented the relevant features of the minimal
UED model in the {\tt CompHEP} event generator \cite{Pukhov:1999gg}.
The MSSM is already available in {\tt CompHEP} since version~41.10. 
All $n=1$ KK modes are incorporated as new
particles, with the proper interactions and
one-loop corrected masses~\cite{Cheng:2002iz}.
The widths can then be readily calculated with {\tt CompHEP}
on a case by case basis and added to the particle table.
Similar to the SM case, the neutral gauge bosons at level~1, 
$Z_1$ and $\gamma_1$, are mixtures of the KK modes of the
hypercharge gauge boson and the neutral $SU(2)_W$ gauge boson.
However, it was shown in~\cite{Cheng:2002ab} that the radiatively corrected
Weinberg angle at level~1 and higher is very small.
For example, $\gamma_1$, which is the LKP in the minimal
UED model, is mostly the KK mode of the hypercharge gauge boson.
For simplicity, in the code we neglect neutral gauge boson mixing
for $n\ge 1$.

In the next section we concentrate on the pair production of level~1
KK muons $e^+e^-\to\mu^+_1\mu^-_1$ and compare it to the analogous
process of smuon pair production in supersymmetry:
$e^+e^-\to\tilde\mu^+\tilde\mu^-$. In UED there are two $n=1$ KK muon Dirac 
fermions: an $SU(2)_W$ doublet $\mu^D_1$ and an $SU(2)_W$ singlet $\mu^S_1$,
both of which contribute in Eq.~(\ref{mu1}) below (see also Fig.~\ref{fig:diagrams_ued}). 
In complete analogy, in supersymmetry, 
there are two smuon eigenstates, $\tilde\mu_L$ and $\tilde\mu_R$, both of which 
contribute in Eq.~(\ref{smuon}). The dominant diagrams in that case 
are shown in Fig.~\ref{fig:diagrams_susy}.
\FIGURE[ht]{
{
\unitlength=1.0 pt
\SetScale{1.0}
\SetWidth{0.7}      
{} 
\allowbreak
\begin{picture}(120,100)(0,0)
\Text(15.0,80.0)[r]{\Blue{$e^+$}}
\Line(40.0,50.0)(20.0,80.0)
\Text(15.0,20.0)[r]{\Blue{$e^-$}}
\Line(40.0,50.0)(20.0,20.0)
\Text(60.0,57.0)[b]{\Blue{$\gamma,Z$}}
\Photon(40.0,50.0)(80.0,50.0){3.0}{6}
\Text(105.0,20.0)[l]{\Red{$\mu^+_1$}}
\Line(80.0,50.0)(100.0,20.0)
\Text(105.0,80.0)[l]{\Red{$\mu^-_1$}}
\Line(80.0,50.0)(100.0,80.0)
\Text(60.0,0.0)[c]{(a)}
\end{picture} \
\begin{picture}(120,100)(0,0)
\Text(15.0,80.0)[r]{\Blue{$e^+$}}
\Line(40.0,50.0)(20.0,80.0)
\Text(15.0,20.0)[r]{\Blue{$e^-$}}
\Line(40.0,50.0)(20.0,20.0)
\Text(60.0,57.0)[b]{\Red{$\gamma_2,Z_2$}}
\Photon(40.0,50.0)(80.0,50.0){3.0}{6}
\Text(105.0,20.0)[l]{\Red{$\mu^+_1$}}
\Line(80.0,50.0)(100.0,20.0)
\Text(105.0,80.0)[l]{\Red{$\mu^-_1$}}
\Line(80.0,50.0)(100.0,80.0)
\GCirc(40,50){3}{0}
\Text(60.0,0.0)[c]{(b)}
\end{picture} \
}
\caption{\label{fig:diagrams_ued}
{\it The dominant Feynman diagrams for KK muon production 
$e^+e^-\rightarrow \mu^+_1\mu^-_1$ in Universal Extra Dimensions. 
The black dot represents a 
KK-number violating boundary interaction \cite{Cheng:2002iz}.}}
}
%
%
%
\FIGURE[ht]{
{
\unitlength=1.0 pt
\SetScale{1.0}
\SetWidth{0.7}      
{} 
\allowbreak
\begin{picture}(120,100)(0,0)
\Text(15.0,80.0)[r]{\Blue{$e^+$}}
\Line(40.0,50.0)(20.0,80.0)
\Text(15.0,20.0)[r]{\Blue{$e^-$}}
\Line(40.0,50.0)(20.0,20.0)
\Text(60.0,57.0)[b]{\Blue{$\gamma,Z$}}
\Photon(40.0,50.0)(80.0,50.0){3.0}{6}
\Text(105.0,20.0)[l]{\Red{$\tilde\mu^+$}}
\DashLine(80.0,50.0)(100.0,20.0){3}
\Text(105.0,80.0)[l]{\Red{$\tilde\mu^-$}}
\DashLine(80.0,50.0)(100.0,80.0){3}
\end{picture} \
}
\caption{\label{fig:diagrams_susy}
{\it The dominant Feynman diagrams for smuon production 
$e^+e^-\rightarrow \tilde\mu^+\tilde\mu^-$ in supersymmetry.}}
}
In principle, there are also diagrams mediated by 
$\gamma_n, Z_n$ for $n=4,6,...$ but they are doubly suppressed -
by the KK-number violating interaction at both vertices and
the KK mass in the propagator -
and here can be safely neglected. However, $\gamma_2$ and $Z_2$ exchange
(Fig.~\ref{fig:diagrams_ued}b)
may lead to resonant production and significant 
enhancement of the cross-section, as well as
interesting phenomenology as discussed below in Section~\ref{sec:photon}.
We have implemented the level~2 neutral gauge bosons $\gamma_2, Z_2$ with their widths, 
including both KK-number preserving and the KK-number violating decays as 
in Ref.~\cite{Cheng:2002ab}.
We consider the final state consisting of two opposite sign muons and 
missing energy. It may arise either from KK muon production in UED

\beq
e^+e^- \to \mu^+_1 \mu^-_1 \to \mu^+ \mu^- \gamma_1 \gamma_1\, ,
\label{mu1}
\eeq

\noindent
with $\gamma_1$ being the LKP, or from smuon pair production in supersymmetry:

\beq
e^+e^- \to \tilde \mu^+ \tilde \mu^- \to \mu^+ \mu^- \tilde \chi^0_1 \tilde \chi^0_1\, ,
\label{smuon}
\eeq

\noindent
where $\tilde\chi^0_1$ is the lightest supersymmetric particle.
We reconstruct the muon energy spectrum and the muon production polar angle, 
aiming at small background from SM processes with minimal biases due to detector 
effects and selection criteria.
The goal is to disentangle KK particle production (\ref{mu1}) in UED
from smuon pair production (\ref{smuon}) in supersymmetry. 
We also determine the masses of the produced particles and test the model 
predictions for the production cross-sections in each case.

We first fix the UED parameters to $R^{-1} =$ 500~GeV, $\Lambda R=$ 20, 
leading to the spectrum given in Table~\ref{tab:uedmasses}.

\begin{table}[ht!]
\begin{center}
\begin{tabular}{|c|c|}
\hline 
Particle  & Mass \\ \hline \hline
$\mu_1^D$  & 515.0 GeV\\
$\mu_1^S$  & 505.4 GeV\\
$\gamma_1$ & 500.9 GeV\\
\hline
\end{tabular}
\end{center} 
\caption{\sl Masses of the KK excitations for the parameters $R^{-1}$ = 500~GeV and
$\Lambda R$ = 20 used in the analysis.} 
\label{tab:uedmasses}
\end{table}

The ISR-corrected signal cross section in UED for the selected final state 
$\mu^+\mu^-\gamma_1\gamma_1$ is 14.4~fb at $\sqrt{s}$ = 3~TeV.
Events have been generated with {\tt CompHEP} and then
reconstructed using a fast simulation based on
parametrized response for a realistic detector at CLIC. In particular, 
the lepton identification efficiency, momentum resolution and polar 
angle coverage are of special relevance to this analysis. We assume that particle 
tracks will be reconstructed through a discrete central tracking system, consisting of 
concentric layers of Si detectors placed in a 4~T solenoidal field. This ensures a 
momentum resolution $\delta p/p^2$ = 4.5$\times 10^{-5}$~GeV$^{-1}$. A forward tracking 
system should provide track reconstruction down to $\simeq 10^{\circ}$. We also account 
for initial state radiation (ISR) and for beamstrahlung effects on the center-of-mass energy.
We assume that muons are identified by their penetration in the instrumented iron 
return yoke of the central coil. A 4~T magnetic field sets an energy cutoff of 
$\simeq$~5~GeV for muon tagging. 

The events from the {\tt CompHEP} generation have been treated with
the {\tt Pythia 6.210} parton shower~\cite{Sjostrand:2001yu} and reconstructed with 
a modified version of the {\tt SimDet 4.0} program~\cite{Pohl:2002vk}. Beamstrahlung 
has been added to the {\tt CompHEP} generation. The luminosity spectrum, obtained by the 
{\tt GuineaPig} beam simulation for the standard CLIC beam parameters at 3~TeV, has been 
parametrised using a modified Yokoya-Chen approximation~\cite{peskin}:

This analysis has backgrounds coming from SM $\mu^+ \mu^- \nu\bar{\nu}$ 
final states, which are mostly due to gauge boson pair production 
$W^+W^- \to \mu^+\mu^- \nu_\mu\bar{\nu}_\mu$, 
$Z^0Z^0 \to \mu^+ \mu^- \nu \bar{\nu}$ and from 
$e^+e^- \to W^+W^- \nu_e \bar{\nu}_e$, 
$e^+e^- \to Z^0Z^0 \nu_e \bar{\nu}_e$, followed by muonic decays. 
The background total cross section is $\simeq$20~fb at $\sqrt{s}$ = 3~TeV. 
In addition to its competitive cross section, this background has leptons produced 
preferentially at small polar angles, therefore biasing the angular distribution. 
In order to reduce this background,
a suitable event selection has been applied. Events have been required
to have two muons, missing energy in excess to 2.5~TeV, transverse energy below 150~GeV 
and event sphericity larger than 0.05. In order to reject the $Z^0Z^0$ background, 
events with di-lepton invariant mass compatible with $M_{Z^0}$ have also been discarded. 
The underlying $\gamma \gamma$ collisions also produce a potential background to this 
analysis in the form of $\gamma \gamma \to \mu^+\mu^-$. This background has been simulated
using the CLIC beam simulation and {\tt Pythia}. Despite its large cross section, it can 
be completely suppressed by a cut on the missing transverse energy $E_T^{missing} >$ 50~GeV.
Finally, in order to remove events with large beamstrahlung, the event sphericity had to be 
smaller than 0.35 and the acolinearity smaller than 0.8. 
These criteria provide a factor $\simeq 30$ background suppression, in the kinematical 
region of interest, while not significantly biasing the lepton momentum distribution. 

\section{Comparison of UED and supersymmetry in $\mu^+\mu^-\emiss$}
\label{sec:contrast}

In order to perform the comparison of UED and MSSM, we adjusted the MSSM parameters 
to get the two smuon masses 
$M_{\tilde\mu_L}$ and $M_{\tilde\mu_R}$ and the lightest neutralino mass 
$M_{\tilde\chi^0_1}$ matching exactly those of the two Kaluza-Klein muons 
$M_{\mu^D_1}$ and $M_{\mu^S_1}$ and of the KK photon $M_{\gamma_1}$ for the 
chosen UED parameters.
It must be stressed that such small mass splitting between the two muon partners is 
typically rather accidental in supersymmetric scenarios. The supersymmetric parameters used are 
given in Table~\ref{tab:susymasses}.

\begin{table}[h!]
\begin{center}
\begin{tabular}{|c|c|}
\hline 
MSSM Parameter  & Value \\
\hline \hline
$\mu$  & 1000 GeV\\
$M_1$  & ~502.65 GeV\\
$M_2$  & 1005.0 GeV\\
$M_{\tilde\mu_L}$ & 512.83 GeV\\
$M_{\tilde\mu_R}$ & 503.63 GeV\\
$\tan \beta$ & 10 \\
\hline
\end{tabular}
\end{center} 
\caption{\sl MSSM parameters for the SUSY study point used in the analysis.
This choice of soft SUSY parameters in {\tt CompHEP} leads to an exact match 
between the corresponding UED and SUSY mass spectra.}
\label{tab:susymasses}
\end{table}

We then simulate both reactions (\ref{mu1}) and (\ref{smuon}) with {\tt CompHEP}
and pass the resulting events through the same simulation and reconstruction.
The ISR-corrected signal cross-section in SUSY for the selected final state
$\mu^+\mu^-\tilde\chi^0_1\tilde\chi^0_1$ is 2.76 fb at $\sqrt{s}$= 3~TeV,
which is about 5 times smaller than in the UED case.

\subsection{Angular distributions and spin measurements}
\label{sec:angles}

In the case of UED, the KK muons are fermions and their angular distribution is given by
\beq
\left(\frac{d\sigma}{d\cos\theta}\right)_{UED} \sim 
1+\frac{E^2_{\mu_1}-M^2_{\mu_1}}{E^2_{\mu_1}+M^2_{\mu_1}}\,\cos^2\theta.
\eeq
Assuming that at CLIC the KK production takes place well above
threshold, the formula simplifies to:
\beq
\left(\frac{d\sigma}{ d\cos\theta}\right)_{UED} \sim 1+\cos^2\theta.
\label{ang_ued}
\eeq
As the supersymmetric muon partners are scalars, the corresponding angular 
distribution is
\beq
\left(\frac{d\sigma}{ d\cos\theta}\right)_{SUSY} \sim 1-\cos^2\theta.
\label{ang_susy}
\eeq

Distributions (\ref{ang_ued}) and (\ref{ang_susy}) are sufficiently distinct to 
discriminate the two cases. However, the polar angles $\theta$ of the original KK-muons 
and smuons are not directly observable and the production polar angles
$\theta_\mu$ of the final state muons are measured instead. But as long as the mass 
differences $M_{\mu_1}-M_{\gamma_1}$ and $M_{\tilde\mu}-M_{\tilde\chi^0_1}$ respectively
remain small, the muon directions are well correlated with those of their parents 
(see Figure~\ref{fig:ang}a).
\FIGURE[ht]{
\epsfig{file=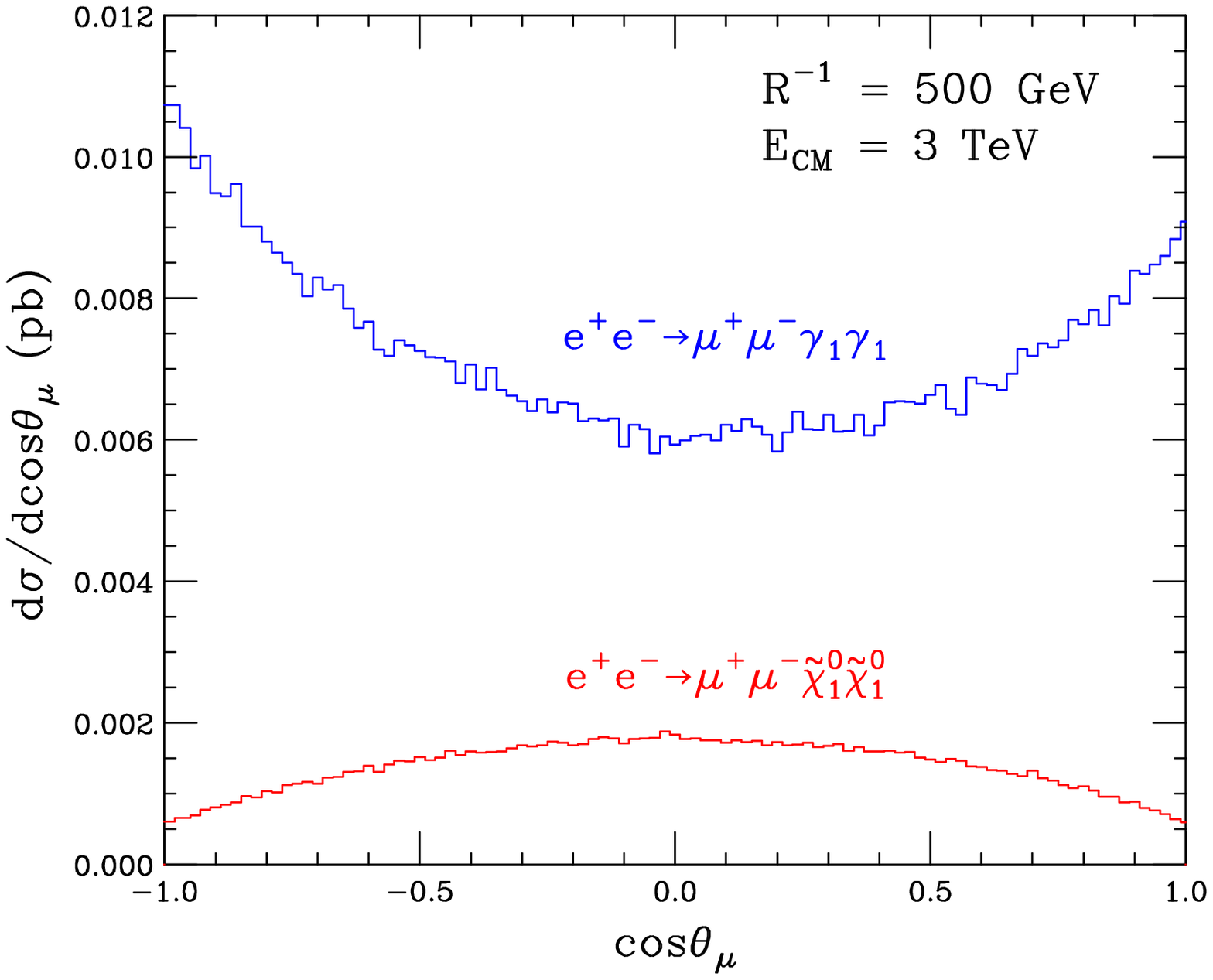,height=5.5cm}
\epsfig{file=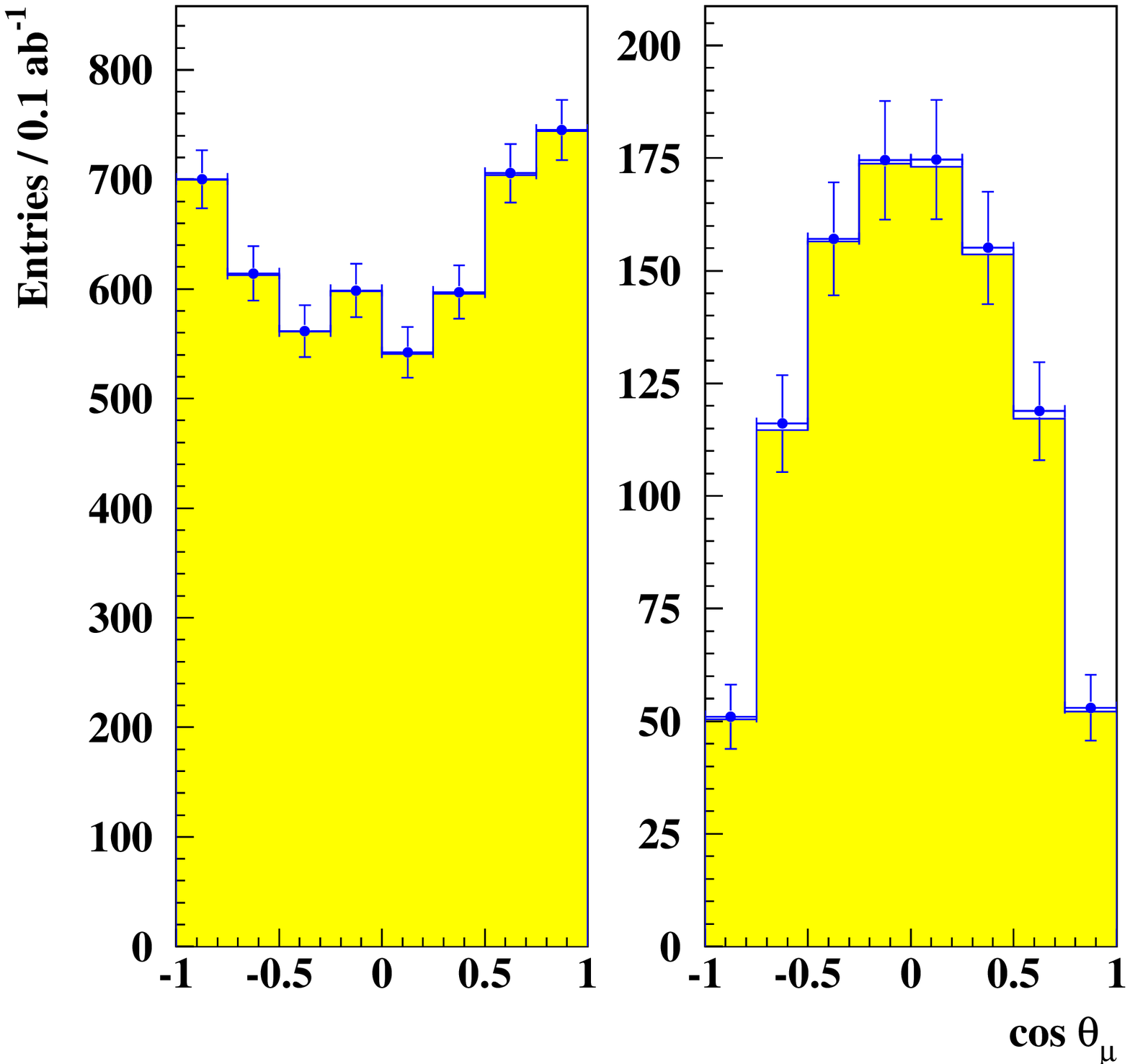,height=6.0cm,width=6.5cm}
\caption{Differential cross-section $d\sigma/d\cos\theta_\mu$ 
for UED (blue, top) and supersymmetry (red, bottom)
as a function of the muon scattering angle $\theta_\mu$.
The figure on the left shows the ISR-corrected theoretical prediction.
The two figures on the right in addition include the effects 
of event selection, beamstrahlung and detector resolution and 
acceptance. The left (right) panel is for the case of UED (supersymmetry).
The data points are the combined signal and background events, while the 
yellow-shaded histogram is the signal only.}
\label{fig:ang}}
In Fig.~\ref{fig:ang}b we show the same comparison after detector simulation and 
including the SM background. The angular distributions are well distinguishable 
also when accounting for these effects. By performing a $\chi^2$ fit to the 
normalised polar angle distribution, the UED scenario considered here could be 
distinguished from the MSSM, on the sole basis of the distribution shape, with 
350~fb$^{-1}$ of data at $\sqrt{s}$ = 3~TeV.

\subsection{Threshold scans}
\label{sec:scan}

At the $e^+e^-$ linear collider, the muon excitation masses can be 
accurately determined through an energy scan of the onset of the pair 
production threshold. This study not only determines the masses, but also 
confirms the particle nature. In fact the cross sections for the UED processes 
rise at threshold $\propto \beta$ while in supersymmetry their threshold onset is 
$\propto \beta^3$, where $\beta$ is the particle velocity. 

\FIGURE[ht]{

\epsfig{file=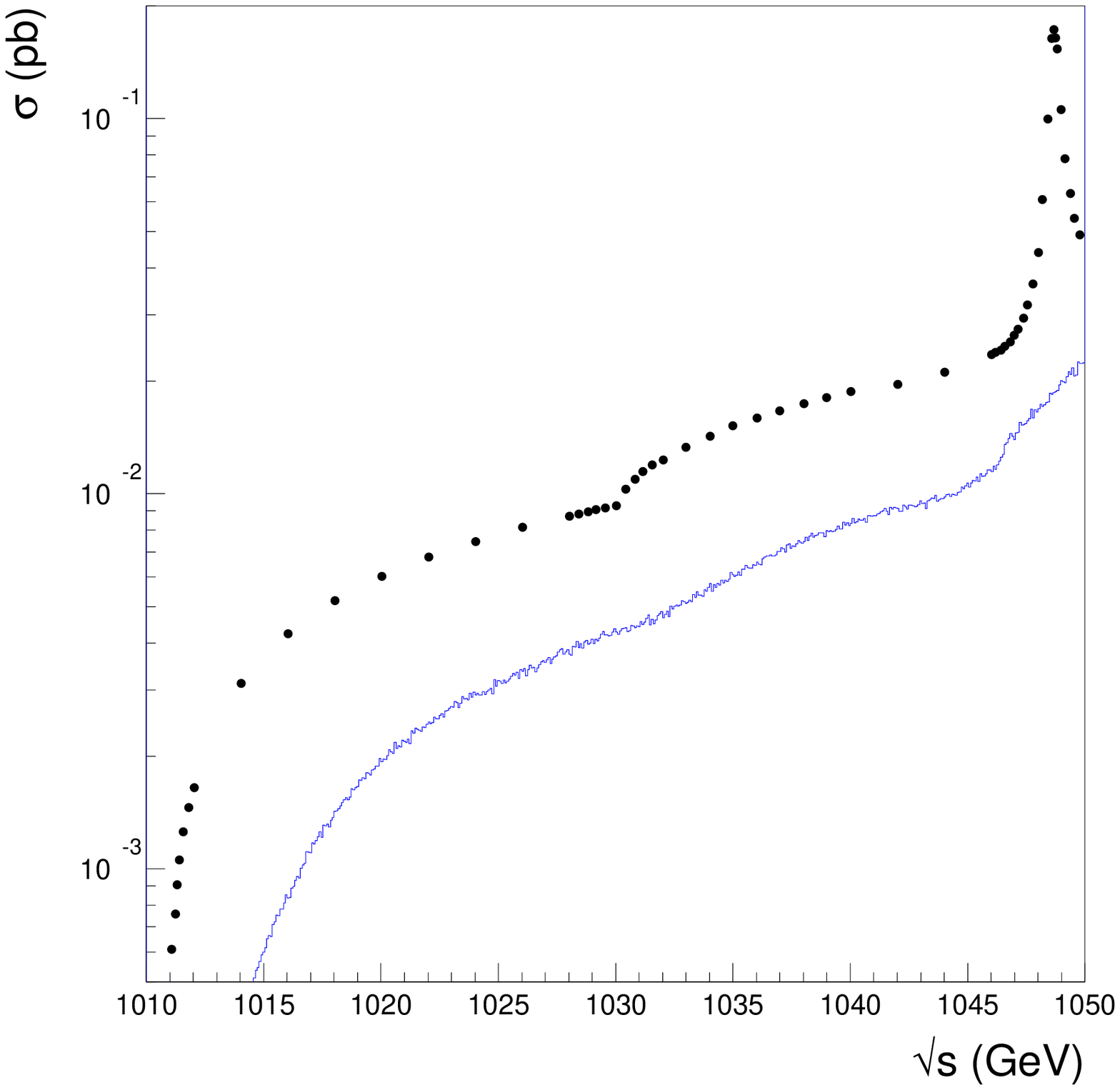,height=7.5cm,width=7.0cm}
\epsfig{file=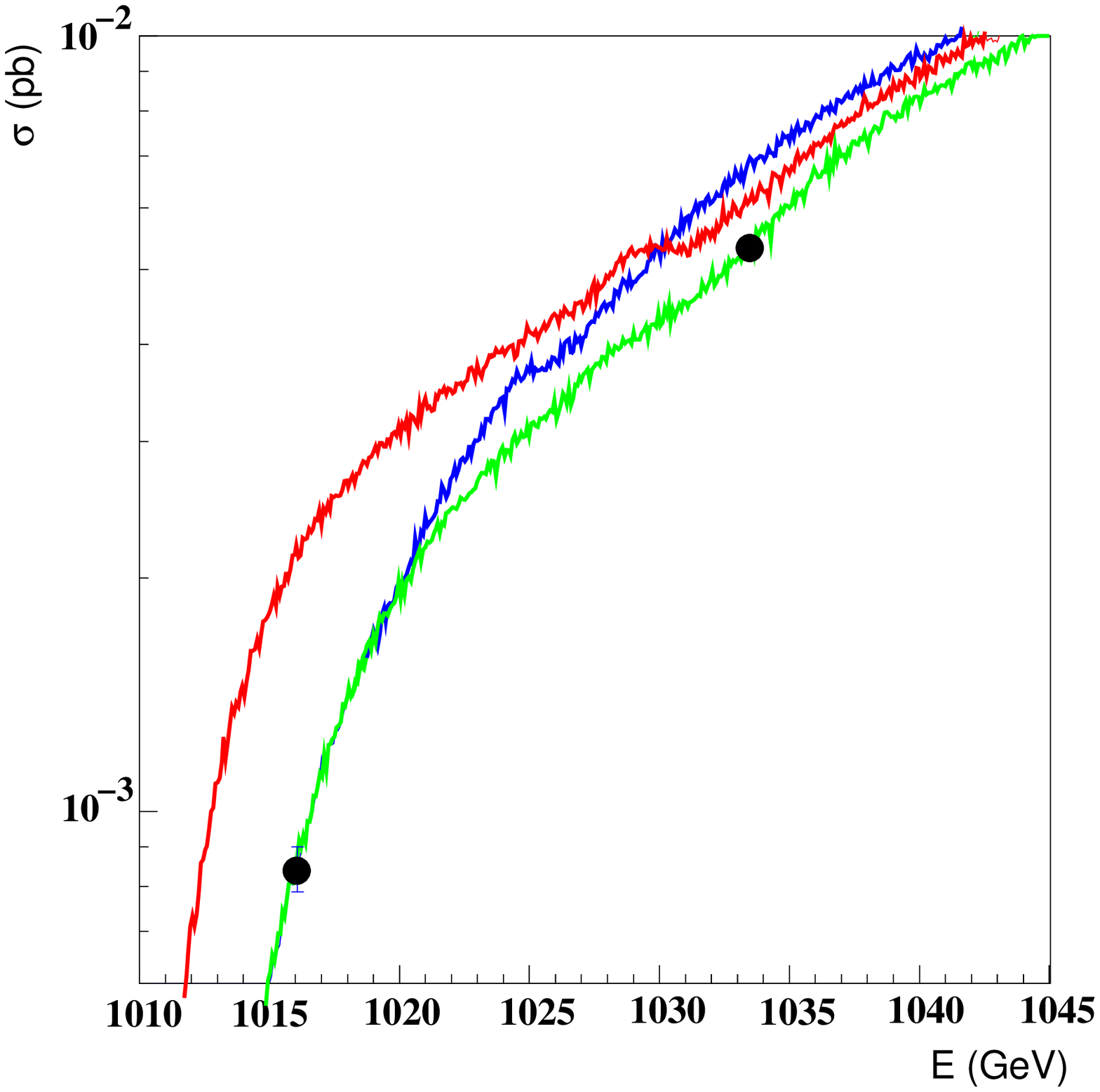,height=7.5cm,width=7.0cm}

\caption{\sl The total $e^+e^- \to \mu^+_1 \mu^-_1 \to \mu^+ \mu^- \gamma_1 \gamma_1$ 
cross section $\sigma$ in pb as a function of the center-of-mass energy $\sqrt{s}$
near threshold. 
Left: the threshold onset with (line, blue) and without (dots) beamstrahlung effects.
Right: a threshold scan at selected points.
The green curve refers to the reference UED parameters while for
the red (blue) curve the mass of $\mu^S_1$ ($\mu^D_1$) has been lowered by
2.5~GeV. The points indicate the expected statistical accuracy for the cross 
section determination at the points of maximum mass sensitivity. 
Effects of the CLIC luminosity spectrum are included.}
\label{fig:scan}}

Since the collision energy can be tuned at properly chosen values, the power rise of 
the cross section can be tested and the masses of the particles involved measured. 
We have studied such threshold scan for the 
$e^+e^- \to \mu^+_1 \mu^-_1 \to \mu^+ \mu^- \gamma_1 \gamma_1$ process at $\sqrt{s}$ = 
1~TeV, for the same parameters as in Table~\ref{tab:uedmasses}. We account for 
the anticipated CLIC centre-of-mass energy spread induced both by the energy spread in 
the CLIC linac and by beam-beam effects during collisions. This been obtained from 
the detailed {\tt GuineaPig} beam simulation and parametrised using the modified Yokoya-Chen 
model~\cite{peskin,Battaglia:2001dg}.  
An optimal scan of a particle pair production threshold consists of just two energy 
points, sharing the total integrated luminosity in equal fractions and chosen at 
energies maximising the sensitivity to the particle widths and masses~\cite{blair}. 
For the UED model scan we have taken three points, one for normalisation and two at the 
maxima of the mass sensitivity (see Figure~\ref{fig:scan}). Inclusion of beamstrahlung 
effects induces a shift of the positions of these maxima towards higher nominal 
$\sqrt{s}$ values~\cite{Battaglia:2002eg}. From the estimated sensitivity 
$d \sigma/dM$ and the cross section accuracy, the masses of the two UED muon excitations
can be determined to $\pm 0.11$~GeV and $\pm 0.23$~GeV for the singlet and the doublet 
states respectively, with a total luminosity of 1~ab$^{-1}$ shared in three points, 
when the particle widths can be disregarded.

\subsection{Production cross section determination}
\label{sec:xsec}

The same analysis can be used to determine the cross section for the process 
$e^+e^- \to \mu^+ \mu^- \emiss$. The SM contribution can be determined 
independently, using anti-tag cuts, and subtracted. Since the cross section for the 
UED process at 3~TeV is about five times larger compared to smuon production in 
supersymmetry, this measurement would reinforce the model identification obtained by the 
spin determination. 
This can be quantified by performing the same $\chi^2$ fit to the 
muon polar production angle discussed above, but now including also the total number 
of selected events. Since the cross section depends on the mass of the pair produced 
particles, we include a systematic uncertainty on the prediction corresponding to a 
$\pm 0.05~\%$ mass uncertainty, which is consistent with the results discussed below.  
At CLIC the absolute luminosity should be measurable to ${\cal{O}}(0.1~\%)$ and the 
average effective collision energy to ${\cal{O}}(0.01~\%)$.

\subsection{Muon energy spectrum and mass measurements}
\label{sec:masses}

The characteristic end-points of the muon energy spectrum are completely determined
by the kinematics of the two-body decay and hence they don't depend on the underlying
framework (SUSY or UED) as long as the masses involved are tuned to be identical.
We show the ISR-corrected expected distributions for the muon energy spectra at
the generator level in Fig.~\ref{fig:emu}a, using the same parameters as in
Fig.~\ref{fig:ang}. As expected, the shape of the $E_\mu$ distribution in the case of
UED coincides with that for MSSM.
\FIGURE[ht]{
\epsfig{file=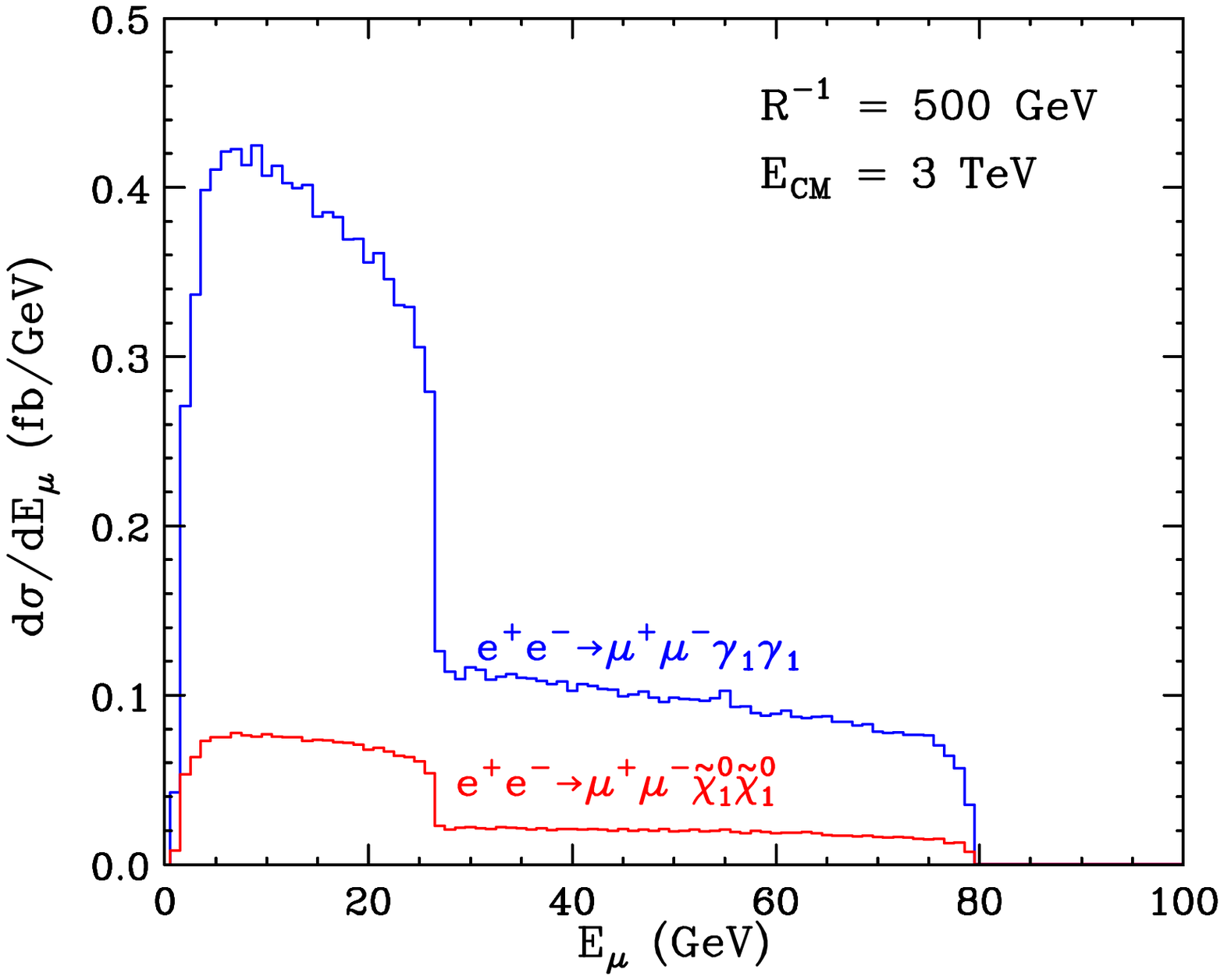,height=5.5cm}
\epsfig{file=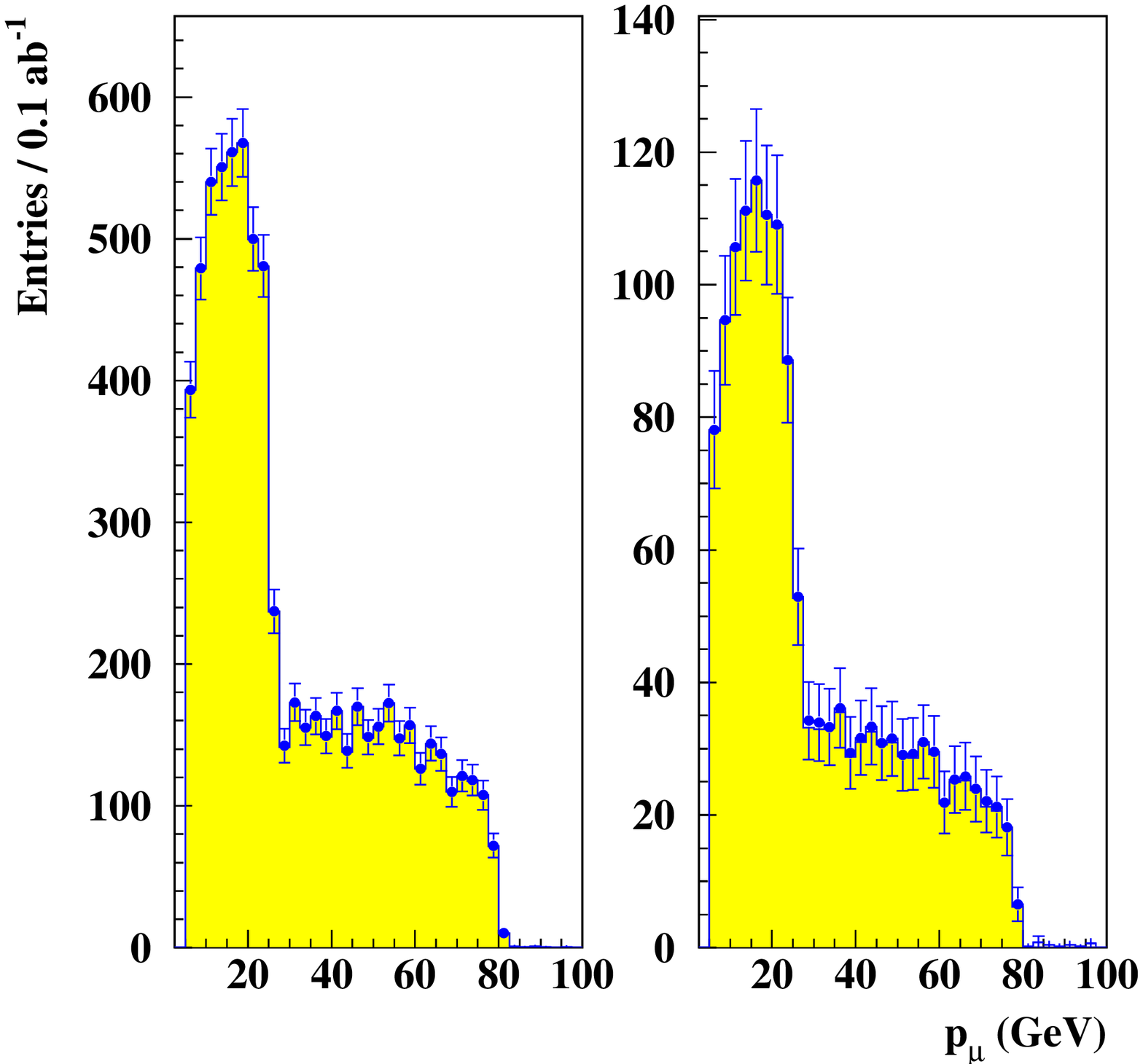,height=6.0cm,width=6.5cm}
\caption{The muon energy spectrum resulting from KK muon production (\ref{mu1})
in UED (blue, top curve) and smuon production (\ref{smuon}) in supersymmetry
(red, bottom curve). The UED and SUSY parameters are chosen as in Fig.~\ref{fig:ang}. 
The plot on the left shows the ISR-corrected distribution,
while that on the right includes in addition the effects of event selection, beamstrahlung 
and detector resolution and acceptance. The data points are the combined signal 
and background events, while the yellow-shaded histogram is the signal only.}
\label{fig:emu}}

The lower, $E_{min}$, and upper, $E_{max}$, endpoints of the muon energy spectrum are 
related to the masses of the particles involved in the decay according to the relation:
\begin{equation}
E_{max/min} = \frac{1}{2} M_{\tilde\mu}
\left(1 - \frac{M^2_{\tilde\chi^0_1}}{M^2_{\tilde\mu}}\right) \gamma (1\pm\beta)
\end{equation}
where $M_{\tilde\mu}$ and $M_{\tilde \chi^0_1}$ are the smuon and LSP masses and 
$\gamma=1/(1-\beta^2)^{1/2}$ with $\beta = \sqrt{1-M^2_{\tilde\mu}/E^2_{beam}}$ is the 
$\tilde\mu$ boost. In the case of the UED the formula is completely analogous with 
$M_{\mu_1}$ replacing $M_{\tilde \mu}$ and $M_{\gamma_1}$ replacing 
$M_{\tilde \chi^0_1}$.

Due to the splitting between the $\tilde \mu_L$ and $\tilde \mu_R$ masses in MSSM and 
that between the $\mu_1^D$ and $\mu_1^S$ masses in UED, in Fig.~\ref{fig:emu}a we see 
the superposition of two box distributions. The left, narrower distribution is due to
$\mu^S_1$ pair production in UED ($\tilde\mu_R$ pair production in supersymmetry).
The underlying, much wider box distribution is due to $\mu^D_1$ pair production in UED 
($\tilde\mu_L$ pair production in supersymmetry). The upper edges are well defined, 
with smearing due to beamstrahlung and, but less importantly, to momentum 
resolution. The lower end of the spectrum has the overlap of the two contributions 
and with the underlying background. Furthermore, since the splitting between the 
masses of the $\mu_1^D$, $\mu_1^S$ and that of $\gamma_1$ is small, the lower end 
of the momentum distribution can be as low as $\cal{O}$(1~GeV) where the lepton 
identification efficiency is cut-off by the solenoidal field bending the lepton
before it reaches the electro-magnetic or the hadron 
calorimeter~\cite{Battaglia:2003wh}.  Nevertheless, 
there is sufficient information in this distribution to extract the mass of 
the $\gamma_1$ particle, using the prior information on the $\mu_1^D$ and 
$\mu_1^S$ masses, obtained by the threshold scan.

In Fig.~\ref{fig:emu}b we show the muon energy distribution after detector simulation.
A one parameter fit gives an uncertainty on the $\gamma_1$ mass of 
$\pm$0.19~(stat.)~$\pm$0.21~(syst)~GeV, where the statistical uncertainty is given 
for 1~ab$^{-1}$ of data and the systematics reflects the effect of the 
uncertainty on the $\mu_1$ masses. The beamstrahlung introduces an additional 
systematics, which depends on the control of the details of the luminosity spectrum.

\subsection{Photon energy spectrum and radiative return to the $Z_2$}
\label{sec:photon}

With the $e^+e^-$ colliding at a fixed centre-of-mass energy above the pair production  
threshold a significant fraction of the KK muon production will proceed through 
radiative return. Since this is mediated by $s$-channel narrow resonances, a sharp peak 
in the photon energy spectrum appears whenever one of the mediating $s$-channel 
particles is on-shell. In case of supersymmetry, only $Z$ and $\gamma$ particles can 
mediate smuon pair production and neither of them can be close to being on-shell.
On the contrary, an interesting feature of the UED scenario is that $\mu_1$ production 
can be mediated by $Z_n$ and $\gamma_n$ KK excitations (for $n$ even)
as shown in Fig.~\ref{fig:diagrams_ued}b. Among these additional contributions, the $Z_2$ 
and $\gamma_2$ exchange diagrams are the most important. Since the decay 
$Z_2\to \mu_1\mu_1$ is allowed by phase space, there will be a sharp peak 
in the photon spectrum, due to a radiative return to the $Z_2$.
The photon peak is at
\beq
E_\gamma = \frac{1}{2}\ E_{CM}\ 
\biggr(1-\frac{M^2_{Z_2}}{ E^2_{CM}} \biggl).
\eeq
On the other hand, $M_{\gamma_2}<2M_{\mu_1}$, so that
the decay $\gamma_2\to\mu_1\mu_1$ is closed, 
and therefore there is no radiative return to $\gamma_2$.
Notice that the level 2 Weinberg angle is very small
\cite{Cheng:2002iz} and therefore $Z_2$ is mostly $W^0_2$-like and
couples predominantly to $\mu^D_1$ and not $\mu^S_1$.

\FIGURE[ht]{
\epsfig{file=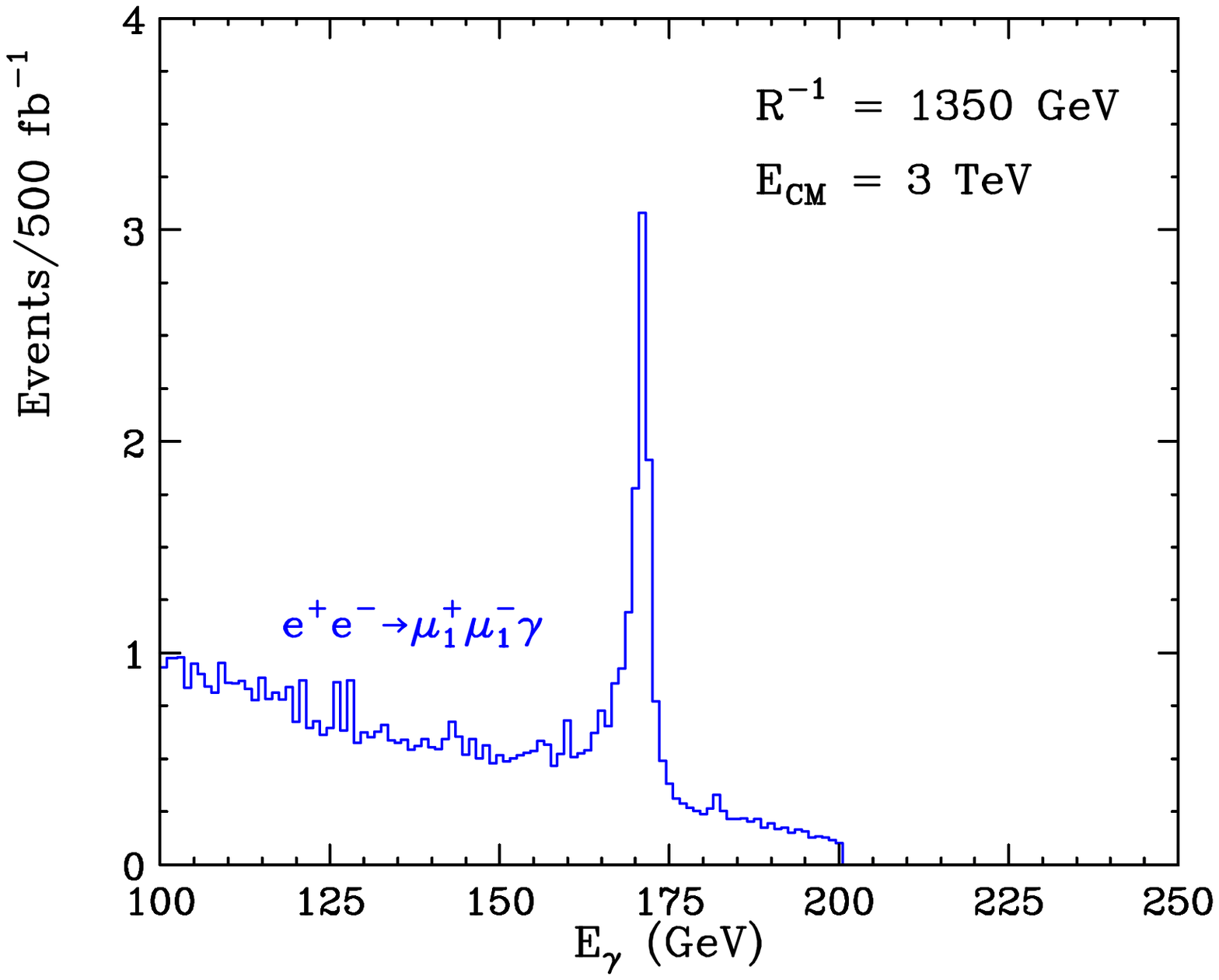,height=5.5cm}
\epsfig{file=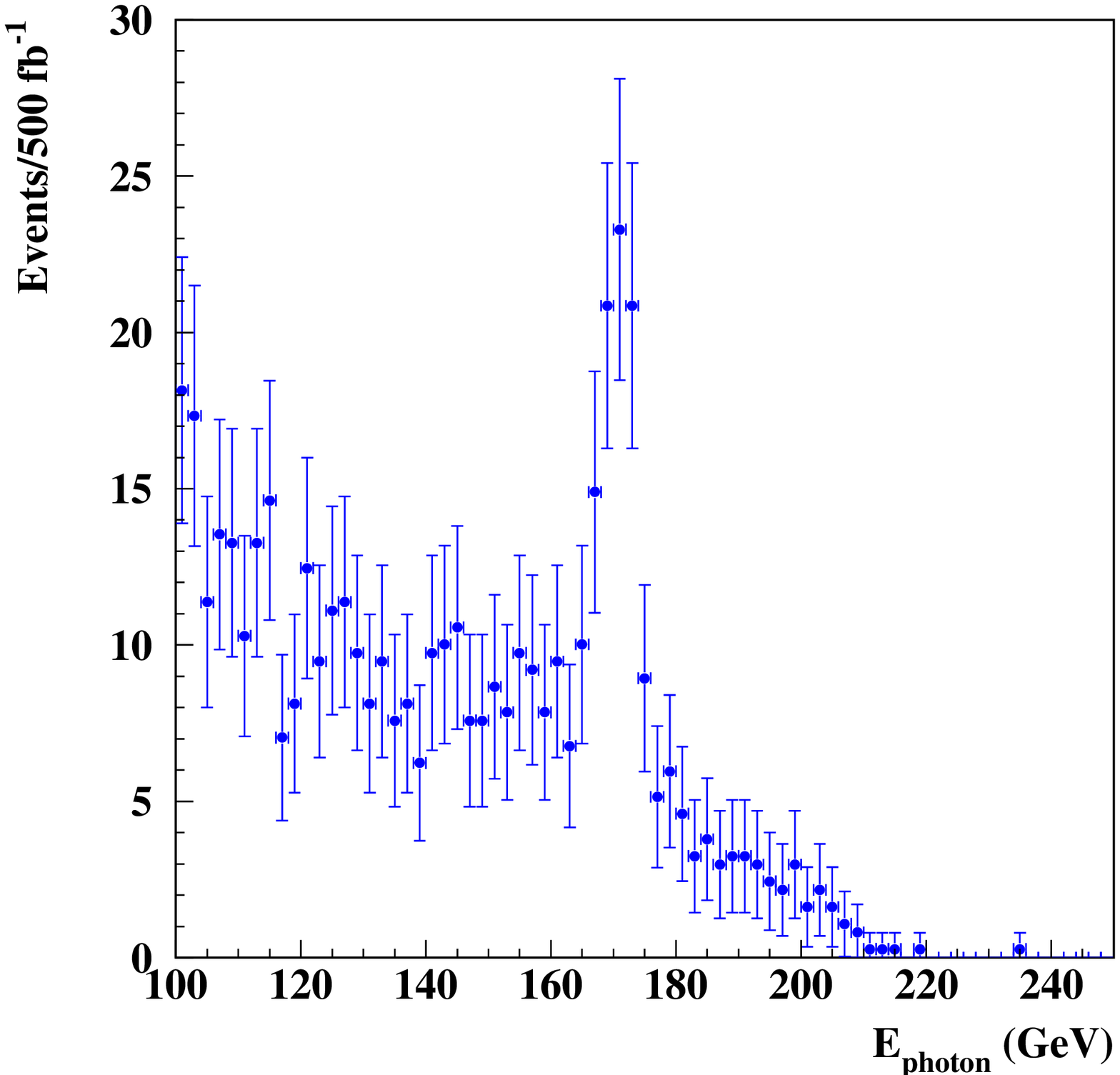,height=5.95cm}
\caption{\sl Photon energy spectrum in $e^+e^-\to \mu^+_1\mu^-_1\gamma$
for $R^{-1}=1350$ GeV, $\Lambda R=20$ and $E_{CM}=3$ TeV before (left) 
and after (right) detector simulation.
The acceptance cuts are $E_\gamma>10$ GeV and $1<\theta_\gamma<179^\circ$.
The mass of the $Z_2$ resonance is 2825 GeV.}
\label{fig:egamma}}

The photon energy spectrum in $e^+e^-\to \mu^+_1\mu^-_1\gamma$
for $R^{-1}=1350$ GeV, $\Lambda R=20$ and
$E_{CM}=3$ TeV is shown in Fig.~\ref{fig:egamma}.
On the left we show the ISR-corrected theoretical prediction
from {\tt CompHEP} while the result on the right in addition includes
detector and beam effects. It is clear that the peak cannot be missed.

\section{Prospects for discovery and discrimination in other final states}
\label{sec:other}

Previously in section~\ref{sec:contrast} we considered the 
$\mu^+\mu^-\emiss$ final state resulting from the pair production of 
level 1 KK muons. However, this is not the only signal which could
be expected in the case of UED. Due to the relative degeneracy of 
the KK particles at each level, the remaining $n=1$ KK modes will be 
produced as well, and will yield observable signatures. In those cases,
the discrimination techniques which we discussed earlier can still be applied,
providing further evidence in favor of one model over the other.
In this section we compute the cross-sections for some of the other main
processes of interest, and discuss how they could be analysed. 

\subsection{Kaluza-Klein leptons}

We first turn to the discussion of the other KK lepton flavors. 
The KK $\tau$-leptons, $\tau_1^\pm$, are also produced in $s$-channel
diagrams only, as in Fig.~\ref{fig:diagrams_ued}, hence the $\tau^+_1\tau^-_1$
production cross-sections are very similar to the $\mu^+_1\mu^-_1$ case.
The final state will be $\tau^+\tau^-\emiss$, and it can be observed 
in several modes, corresponding to the different options for the $\tau$ decays.
However, due to the lower statistics and the inferior jet energy resolution,
none of the resulting channels can compete with the discriminating power of the 
$\mu^+_1\mu^-_1\emiss$ final state discussed in the previous section.

The case of KK electrons is more interesting, as it contains a new twist.
The production of KK electrons can also proceed through the $t$-channel diagram 
shown in Fig.~\ref{fig:diagrams_ued_ee}c.
\FIGURE[ht]{
{
\unitlength=1.0 pt
\SetScale{1.0}
\SetWidth{0.7}      
{} 
\allowbreak
\begin{picture}(120,100)(0,0)
\Text(15.0,80.0)[r]{\Blue{$e^+$}}
\Line(40.0,50.0)(20.0,80.0)
\Text(15.0,20.0)[r]{\Blue{$e^-$}}
\Line(40.0,50.0)(20.0,20.0)
\Text(60.0,57.0)[b]{\Blue{$\gamma,Z$}}
\Photon(40.0,50.0)(80.0,50.0){3.0}{6}
\Text(105.0,20.0)[l]{\Red{$e^-_1$}}
\Line(80.0,50.0)(100.0,20.0)
\Text(105.0,80.0)[l]{\Red{$e^+_1$}}
\Line(80.0,50.0)(100.0,80.0)
\Text(60.0,0.0)[c]{(a)}
\end{picture} \
\begin{picture}(120,100)(0,0)
\Text(15.0,80.0)[r]{\Blue{$e^+$}}
\Line(40.0,50.0)(20.0,80.0)
\Text(15.0,20.0)[r]{\Blue{$e^-$}}
\Line(40.0,50.0)(20.0,20.0)
\Text(60.0,57.0)[b]{\Red{$\gamma_2,Z_2$}}
\Photon(40.0,50.0)(80.0,50.0){3.0}{6}
\Text(105.0,20.0)[l]{\Red{$e^-_1$}}
\Line(80.0,50.0)(100.0,20.0)
\Text(105.0,80.0)[l]{\Red{$e^+_1$}}
\Line(80.0,50.0)(100.0,80.0)
\GCirc(40,50){3}{0}
\Text(60.0,0.0)[c]{(b)}
\end{picture} \
\begin{picture}(120,100)(0,0)
\Text(15.0,80.0)[r]{\Blue{$e^+$}}
\Line(60.0,80.0)(20.0,80.0)
\Text(15.0,20.0)[r]{\Blue{$e^-$}}
\Line(60.0,20.0)(20.0,20.0)
\Text(67.0,50.0)[l]{\Red{$\gamma_1,Z_1$}}
\Photon(60.0,80.0)(60.0,20.0){3.0}{7}
\Text(105.0,20.0)[l]{\Red{$e^-_1$}}
\Line(60.0,20.0)(100.0,20.0)
\Text(105.0,80.0)[l]{\Red{$e^+_1$}}
\Line(60.0,80.0)(100.0,80.0)
\Text(60.0,0.0)[c]{(c)}
\end{picture} \
}
\caption{\label{fig:diagrams_ued_ee}
{\it The same as Fig.~\ref{fig:diagrams_ued}, but for KK electron production 
$e^+e^-\rightarrow e^+_1 e^-_1$.}}
}
As a result, the production cross-sections for KK electrons can be much higher
than for KK muons. We illustrate this in Fig.~\ref{fig:KKleptons}, where 
we show separately the cross-sections for $SU(2)_W$ doublets 
(solid lines) and $SU(2)_W$ singlets (dotted lines), as a function of $R^{-1}$.
(For the numerical results throughout section~\ref{sec:other}, we always fix
$\Lambda R=20$.) 
At low masses (i.e. low $R^{-1}$) the $e^+_1e^-_1$ cross-sections can 
be up to two orders of magnitude larger, compared to the case of 
$\mu^+_1\mu^-_1$. Another interesting feature is the resonant enhancement of the 
cross-section for $R^{-1}\sim 1450$ GeV, which is present
in either case ($e$ or $\mu$) for the $SU(2)_W$ doublets (solid lines), 
but not the $SU(2)_W$ singlets (dotted lines). The feature is due to the on-shell 
production of the level 2 $Z_2$ KK gauge boson, which can then decay
into a pair of level 1 KK leptons (see diagram (b) in Figs.~\ref{fig:diagrams_ued}
and \ref{fig:diagrams_ued_ee}). Since the Weinberg angle at the higher ($n>0$)
KK levels is tiny \cite{Cheng:2002iz}, $Z_2$ is predominantly an $SU(2)_W$ gauge boson and
hence does not couple to the $SU(2)_W$ singlet fermions, which explains the 
absence of a similar peak in the $e^S_1$ and $\mu^S_1$ cross-sections\footnote{One 
might have expected a second peak closeby due to $\gamma_2$ resonant production, 
but in the minimal UED model the spectrum is such that the decays of 
$\gamma_2$ to level 1 fermions are all closed.}.
\FIGURE[ht]{
\epsfig{file=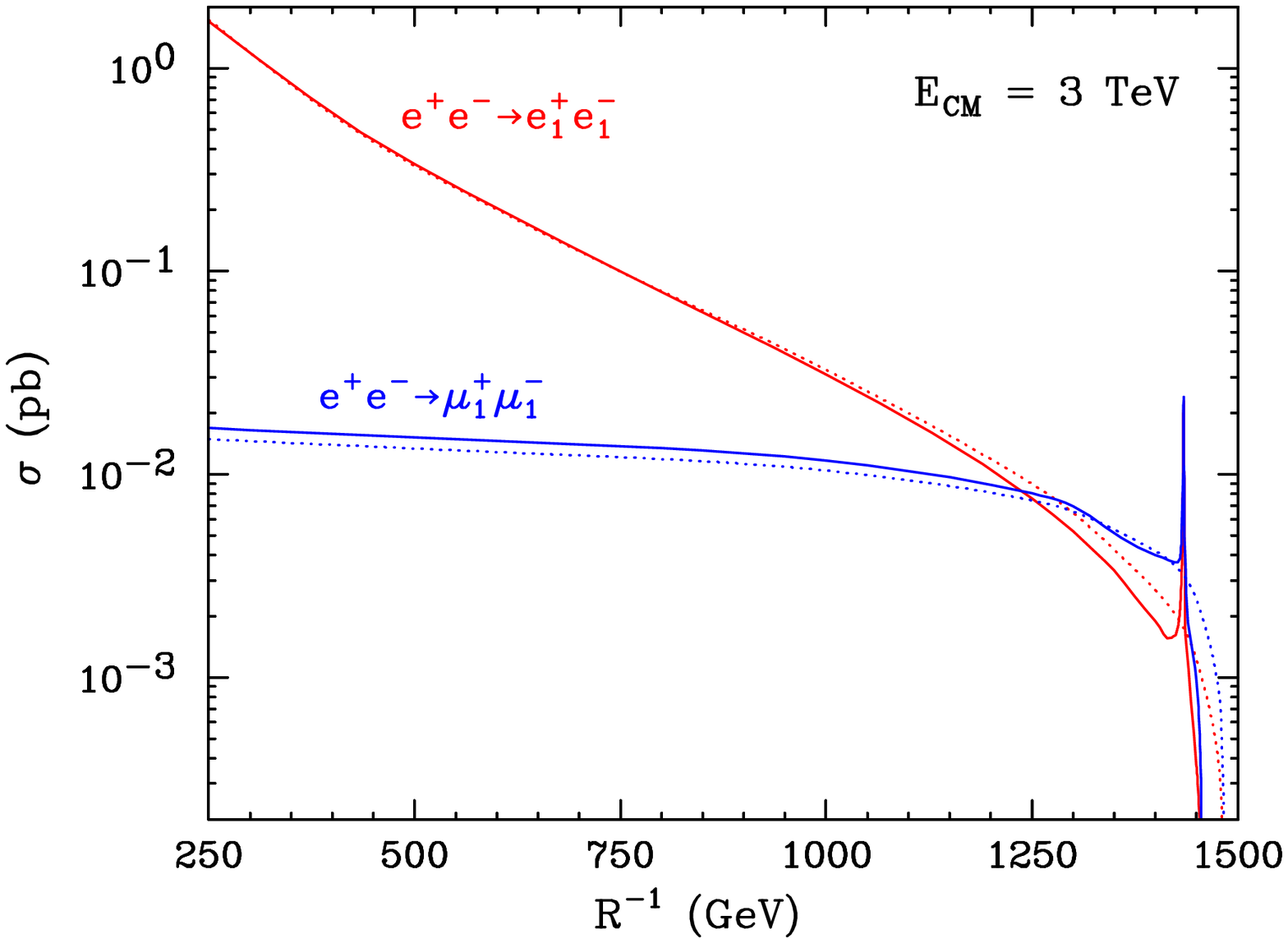,height=7.0cm}
\caption{\sl ISR-corrected production cross-sections of level 1 KK leptons 
($e_1$ in red, $\mu_1$ in blue) at CLIC, as a function of $R^{-1}$. 
Solid (dotted) lines correspond to $SU(2)_W$ doublets (singlets).}
\label{fig:KKleptons}}

Because of the higher production rates, the $e^+e^-\emiss$ event sample 
will be much larger and have better statistics than $\mu^+\mu^-\emiss$.
The $e^+e^-\emiss$ final state has been recently advertised as a discriminator
between UED and supersymmetry in~\cite{Bhattacharyya:2005vm}. 
However, the additional $t$-channel diagram (Fig.~\ref{fig:diagrams_ued_ee}c) 
has the effect of not only enhancing the overall cross-section, 
but also distorting the differential angular distributions 
discussed previously in Section~\ref{sec:angles}, and
creating a forward peak, which causes the cases of UED and supersymmetry 
to look very much alike. We show the resulting angular distributions
of the final state electrons in Fig.~\ref{fig:ang_ee}. For proper comparison, 
we follow the same procedure as before: we choose the UED spectrum 
for $R^{-1}=500$ GeV, which yields KK electron masses as in 
Table~\ref{tab:uedmasses}. We then choose a supersymmetric spectrum with
selectron mass parameters as in Table~\ref{tab:susymasses}. This guarantees 
matching mass spectra in the two cases (UED and supersymmetry) so that any
differences in the angular distributions should be attributed to the
different spins. 
\FIGURE[ht]{
\epsfig{file=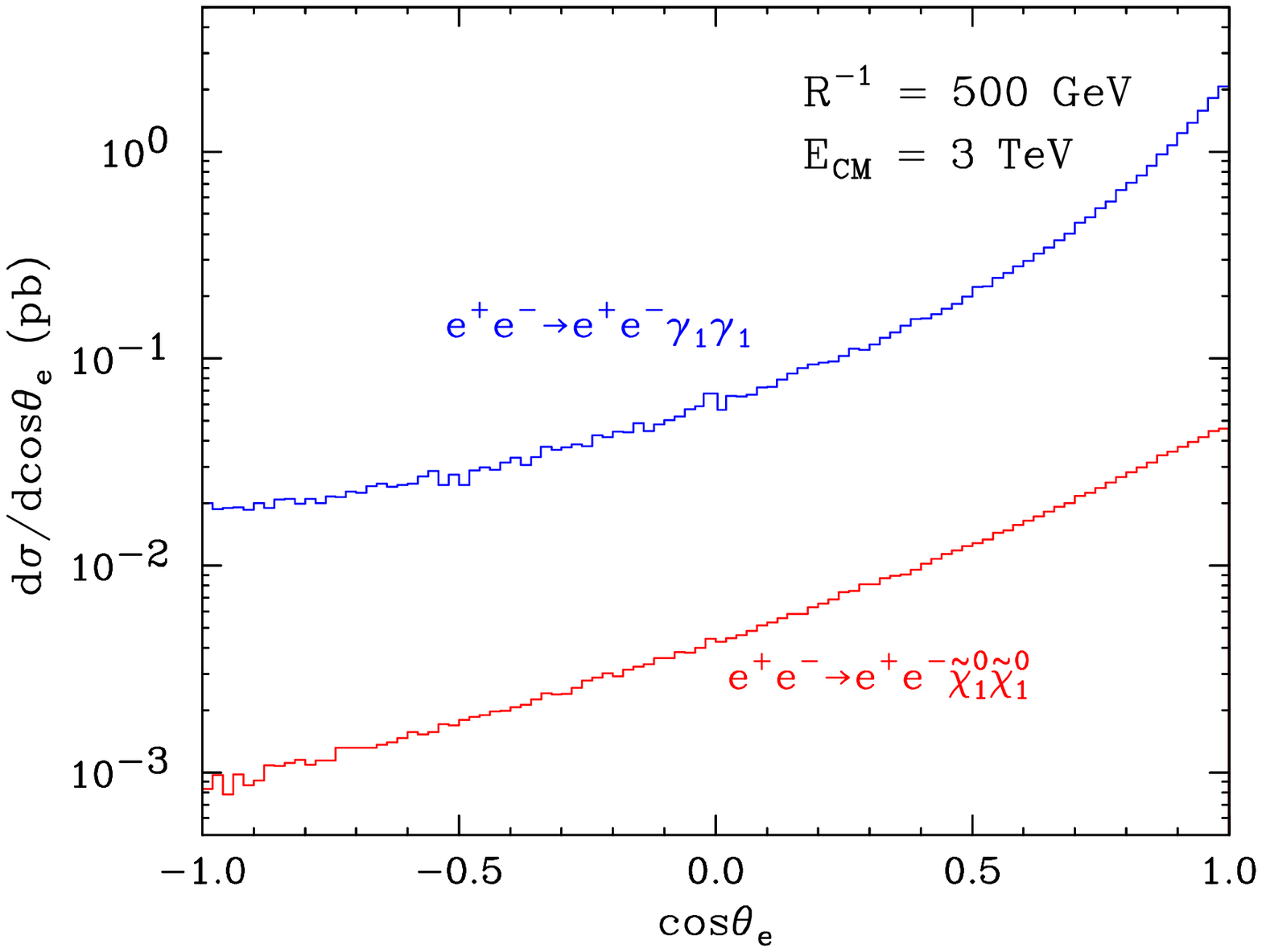,height=7.0cm}
\caption{\sl The same as Fig.~\ref{fig:ang} (left panel), but for 
KK electron production $e^+e^-\rightarrow e^+_1 e^-_1$, with
$\theta_e$ being the electron scattering angle.}
\label{fig:ang_ee}}

Unlike Fig.~\ref{fig:ang}, where the underlying shapes of the 
angular distributions were very distinctive (see eqs.~(\ref{ang_ued}) 
and (\ref{ang_susy})), the main effect in Fig.~\ref{fig:ang_ee} 
is the uniform enhancement of the forward scattering cross-section, 
which tends to wash out the spin correlations exhibited in Fig.~\ref{fig:ang}.

\subsection{Kaluza-Klein quarks}

Level 1 KK quarks will be produced in $s$-channel via diagrams similar 
to those exhibited in Fig.~\ref{fig:diagrams_ued}. The corresponding production 
cross-sections are shown in Fig.~\ref{fig:KKquarks}, as a function of
$R^{-1}$. We show separately the cases of the $SU(2)_W$ doublets $u^D_1$ and $d^D_1$
and the $SU(2)_W$ singlets $u^S_1$ and $d^S_1$. In the minimal UED model,
the KK fermion doublets are somewhat heavier than the KK fermion singlets 
\cite{Cheng:2002iz}, so naturally, the production cross-sections for
$u^D_1$ and $d^D_1$ cut off at a smaller value of $R^{-1}$.
Since singlet production is only mediated by $U(1)$ hypercharge interactions,
the singlet production cross-sections tend to be smaller. We notice that 
$u^S_1\bar{u}^S_1$ is larger by a factor of $2^2$ compared to $d^S_1\bar{d}^S_1$, 
in accordance with the usual quark hypercharge assignments.
\FIGURE[ht]{
\epsfig{file=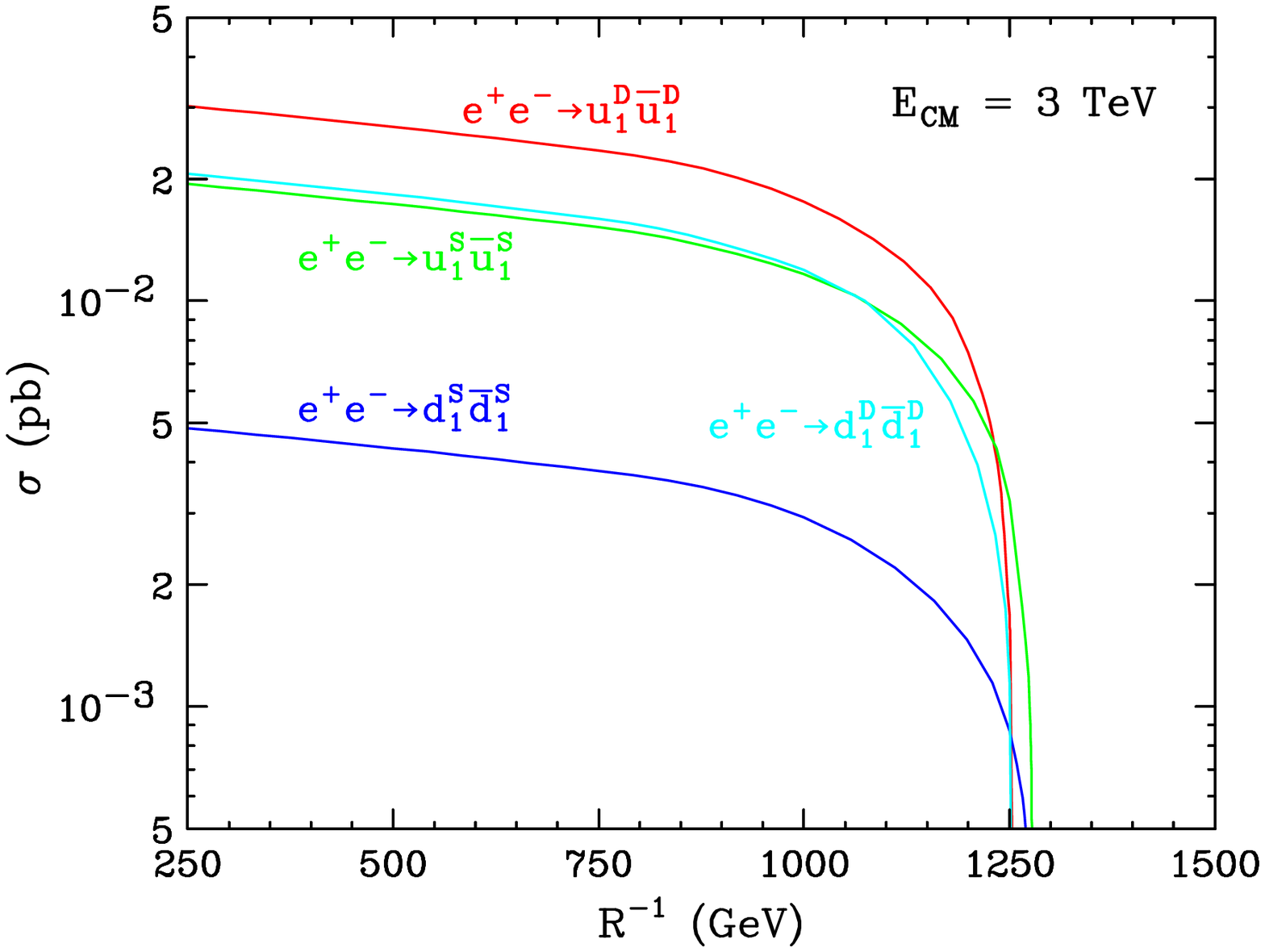,height=7.0cm}
\caption{\sl ISR-corrected production cross-sections of level 1 KK quarks 
at CLIC, as a function of $R^{-1}$.}
\label{fig:KKquarks}}

The observable signals will be different in the case of $SU(2)_W$ 
doublets and $SU(2)_W$ singlets. The singlets, $u^S_1$ and $d^S_1$,
decay directly to the LKP $\gamma_1$, and the corresponding signature 
will be 2 jets and missing energy. The jet angular distribution will 
again be indicative of the KK quark spin, and can be used to discriminate 
against (right-handed) squark production in supersymmetry, following the
procedure outlined in section~\ref{sec:angles}. The jet energy distribution
will again exhibit endpoints, which will in principle allow for the 
mass measurements discussed in section~\ref{sec:masses}. A threshold 
scan of the cross-section will provide further evidence of the 
particle spins (see section~\ref{sec:scan}). The only major difference 
with respect to the $\mu^+\mu^-\emiss$ final state discussed in 
section~\ref{sec:contrast}, is the absence of the monochromatic photon 
signal from section~\ref{sec:photon}, since $Z_2$ is too light to decay to 
KK quarks. In spite of the many similarities to the dimuon final state 
considered in section~\ref{sec:contrast}, notice that jet angular and energy
measurements are not as clean and therefore the lepton (muon or electron) 
final states would still provide the most convincing evidence for discrimination.

The signatures of the $SU(2)_W$ doublet quarks are richer -- 
both $u^D_1$ and $d^D_1$ predominantly decay to $Z_1$ and $W^\pm_1$
which in turn decay to leptons and the LKP~\cite{Cheng:2002ab}.
The analogous process in supersymmetry would be left-handed squark production with
subsequent decays to $\tilde\chi^0_2$ or $\tilde\chi^\pm_1$, 
which in turn decay to $\tilde\ell_L$ and $\tilde \chi^0_1$.
In principle, the spin information will still be encoded in the angular 
distributions of the final state particles. However, the analysis is much more involved, 
due to the complexity of the signature, and possibly the additional missing energy 
from any neutrinos.

\subsection{Kaluza-Klein gauge bosons}

The ISR-corrected production cross-sections for level 1 electroweak\footnote{The level 1 
KK gluon, of course, has no tree-level couplings to $e^+e^-$.} KK gauge bosons 
($W^\pm_1$, $Z_1$ and $\gamma_1$) at a 3 TeV $e^+e^-$ collider are shown 
in Fig.~\ref{fig:KKV1}, as a function of $R^{-1}$. The three relevant processes
are $W^+_1W^-_1$, $Z_1Z_1$ and $Z_1\gamma_1$ ($\gamma_1\gamma_1$ is unobservable).
In each case, the production can be mediated by a $t$-channel exchange of a 
level 1 KK lepton, while for $W^+_1W^-_1$ there are additional $s$-channel
diagrams with $\gamma$, $Z$, $\gamma_2$ and $Z_2$.
$Z_1$ and $W^\pm_1$ are almost degenerate~\cite{Cheng:2002iz}, thus their 
cross-sections cut off at around the same point. The analogous processes in 
supersymmetry would be the pair production of gaugino-like charginos and neutralinos.
The final states will always involve leptons and missing energy, since 
$W^\pm_1$ and $Z_1$ do not decay to KK quarks. 
\FIGURE[ht]{
\epsfig{file=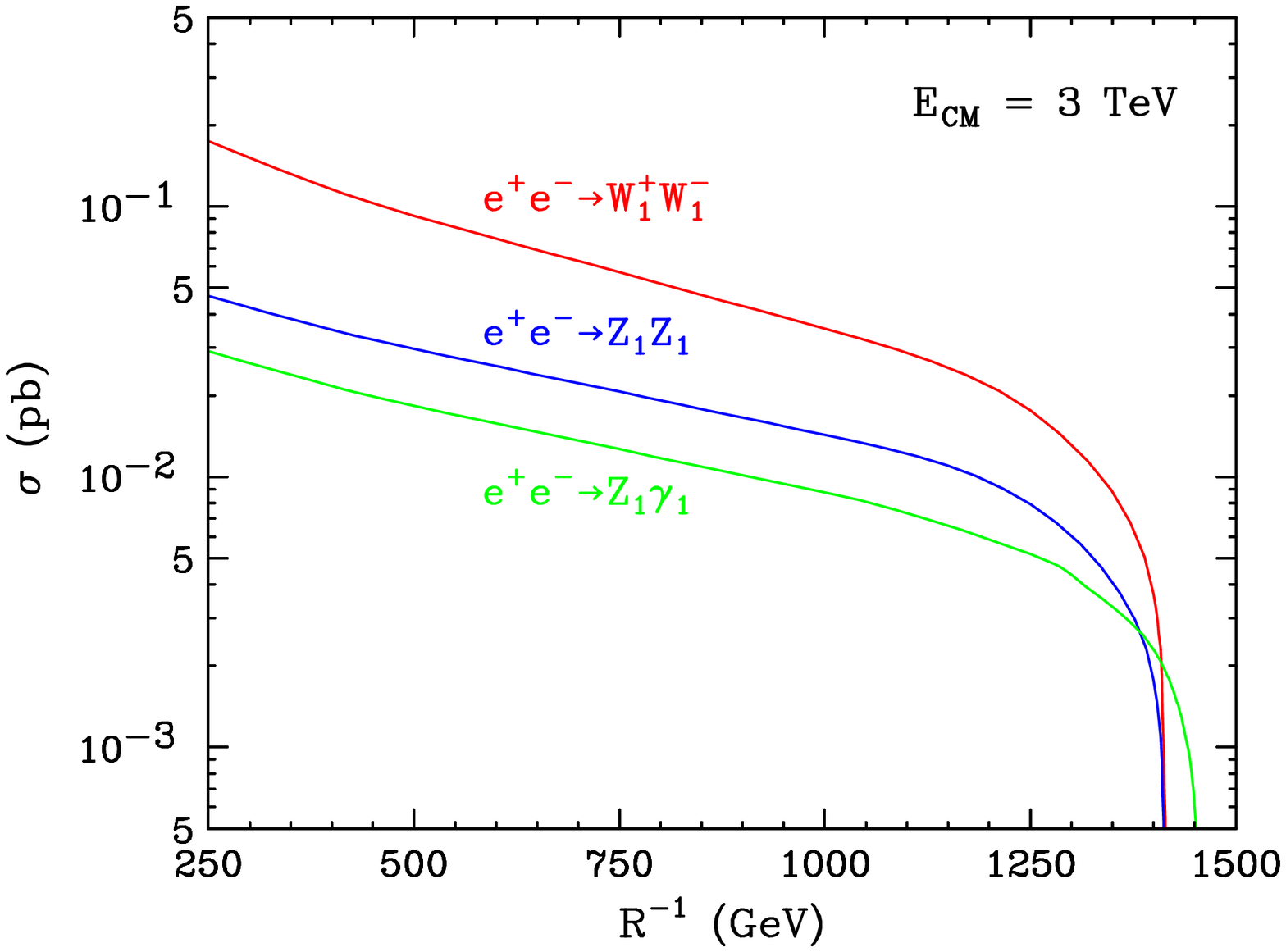,height=7.0cm}
\caption{\sl ISR-corrected production cross-sections of level 1 KK gauge bosons
at CLIC, as a function of $R^{-1}$.}
\label{fig:KKV1}}

In conclusion of this section, for completeness we also discuss the
possibility of observing the higher level KK particles and in particular
those at level 2. For small enough $R^{-1}$, level 2 KK modes 
are kinematically accessible at CLIC. Once produced, they will 
in general decay to level 1 particles and thus contribute to the 
inclusive production of level 1 KK modes. Uncovering the presence of the
level 2 signal in that case seems challenging, but not impossible.

We choose to concentrate on the case of the level 2 KK gauge bosons 
($V_2$), which are somewhat special in the sense that they can decay
directly to SM fermions through KK number violating interactions. Thus
they can be easily observed as dijet or dilepton resonances.
In principle, there are two types of production mechanisms 
for level 2 gauge bosons. The first is single production $e^+e^-\to V_2$,
which can only proceed through KK number violating (loop suppressed) couplings.
The second mechanism is $e^+e^-\to V_2V_2$ pair production 
which is predominantly due to KK number conserving (tree-level) couplings.
In Fig.~\ref{fig:KKV2} we show the corresponding cross-sections 
for the case of the neutral level 2 gauge bosons, as a function of $R^{-1}$,
For low values of $R^{-1}$, pair production dominates, but 
as the level 2 gauge boson masses increase and approach 
$E_{CM}$, single production becomes resonantly enhanced.
Thus the first indication of the presence of the level 2 
particles may come from pair production events, but once the 
mass of the dijet or dilepton resonance is known, the collider 
energy can be tuned to enhance the cross-section and study the
$V_2$ resonance properties in great detail.
\FIGURE[ht]{
\epsfig{file=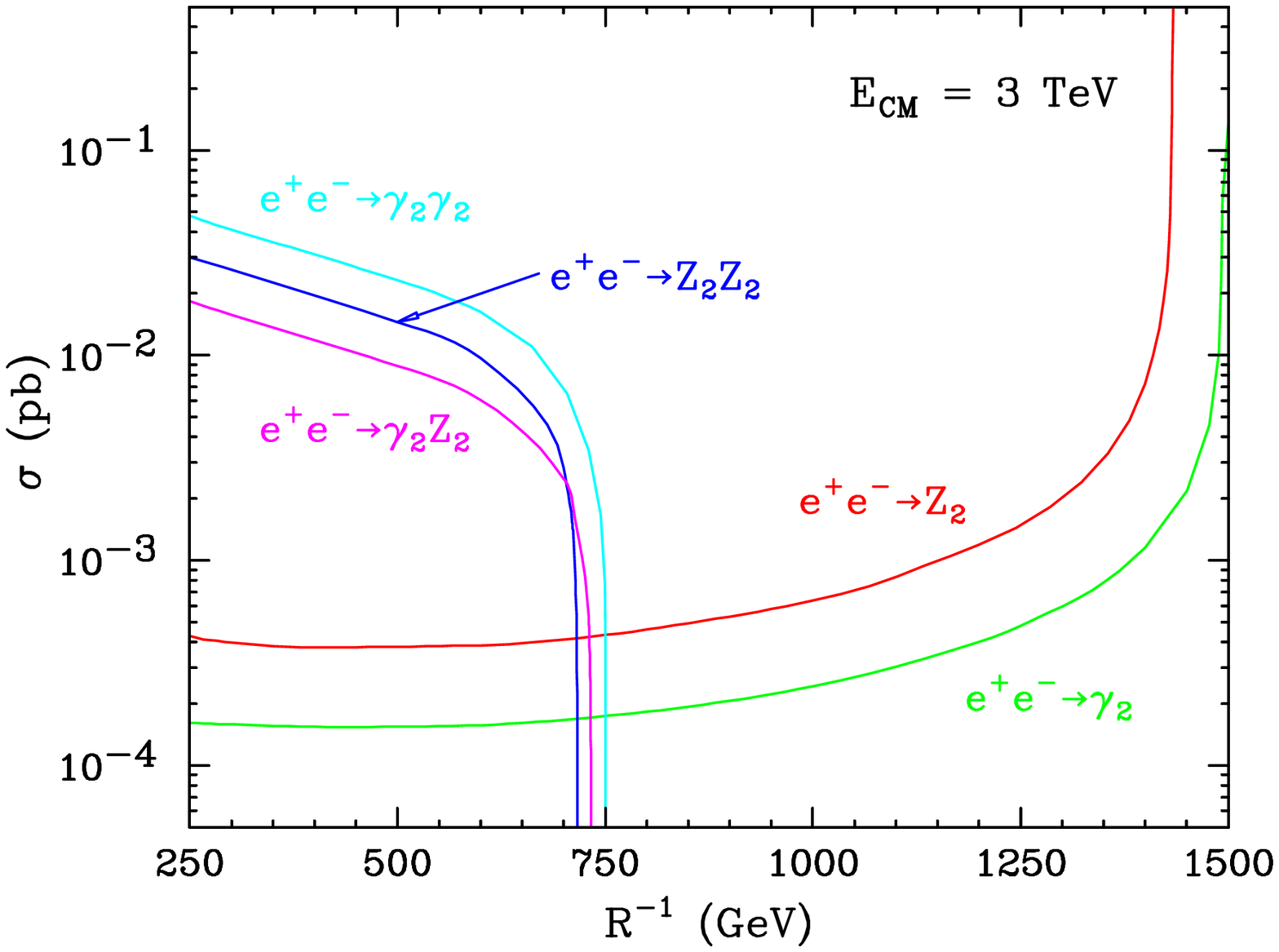,height=7.0cm}
\caption{\sl ISR-corrected production cross-sections of level 2 KK gauge bosons
at CLIC, as a function of $R^{-1}$.}
\label{fig:KKV2}}

\section{Conclusions}
\label{sec:conclusions}

Supersymmetry and Universal Extra Dimensions are two appealing examples
of new physics at the TeV scale, as they address some of the theoretical
puzzles of the SM. They also provide a dark matter 
candidate which, for properly chosen theory parameters, is consistent with present 
cosmology data. Both theories predict a host of new particles, partners of 
the known SM particles. If either one is realised in nature, the LHC is expected to 
observe signals of these new particles. However, in order to clearly identify 
the nature of the new physics, one may need to contrast the UED and 
supersymmetric hypotheses at a multi-TeV $e^+e^-$ linear collider such as 
CLIC\footnote{Similar studies can also be done at the ILC provided the level 1
KK particles are within its kinematic reach. Since precision data tends to
indicate the bound $R^{-1}\ge 250$ GeV for the case of 1 extra dimension,
one would need an ILC center-of-mass energy above 500 GeV in order 
to pair-produce the lowest lying KK states of the minimal UED model.}. 
In this paper we studied in detail the process of pair production of muon 
partners in the two theories, KK-muons and smuons respectively.
We used the polar production angle to distinguish 
the nature of the particle partners, based on their spin. 
The same analysis could be applied for the case of other 
KK fermions, as discussed in section~\ref{sec:other}. 

We have also studied the accuracy of CLIC in determining the masses of the new particles 
involved both through the study of the energy distribution of final state muons and 
threshold scans. An accuracy of better than 0.1\% can be obtained with 1~ab$^{-1}$ of 
integrated luminosity. Once the masses of the partners are known, the measurement 
of the total cross section serves as an additional cross-check on the
hypothesized spin and couplings of the new particles. A peculiar feature of UED, 
which is not present in supersymmetry, is the sharp peak in the ISR photon energy 
spectrum due to a radiative return to the KK partner of the $Z$.

The clean final states and the control over the centre-of-mass energy at the CLIC 
multi-TeV collider allows one to unambiguously identify the nature of the new physics
signals which might be emerging at the LHC already by the end of this decade.

\bigskip

\acknowledgments
AD is supported by the US Department of Energy and the 
Michigan Center for Theoretical Physics.
The work of AD, KK and KM is supported in part by 
a US Department of Energy Outstanding Junior Investigator award 
under grant DE-FG02-97ER41209.
AD, KK and KM acknowledge helpful correspondence with A.~Pukhov.

\listoftables           
\listoffigures          



\begin{thebibliography}{999}

\bibitem{Assmann:2000hg}
The CLIC Study Group, (G.~Guignard ed.), 
{\it A 3-TeV e+ e- linear collider based on CLIC technology},
CERN-2000-008.

\bibitem{Group:2004sz}
C.~P.~W.~Group {\it et al.},
``Physics at the CLIC multi-TeV linear collider,''
arXiv:hep-ph/0412251.

\bibitem{Appelquist:2000nn}
T.~Appelquist, H.~C.~Cheng and B.~A.~Dobrescu,
Phys.\ Rev.\ D {\bf 64}, 035002 (2001)
[arXiv:hep-ph/0012100].

\bibitem{Cheng:2002ab}
H.~C.~Cheng, K.~T.~Matchev and M.~Schmaltz,
Phys.\ Rev.\ D {\bf 66}, 056006 (2002)
[arXiv:hep-ph/0205314].

\bibitem{Rizzo:2001sd}
T.~G.~Rizzo,
Phys.\ Rev.\ D {\bf 64}, 095010 (2001)
[arXiv:hep-ph/0106336].

\bibitem{Macesanu:2002db}
C.~Macesanu, C.~D.~McMullen and S.~Nandi,
Phys.\ Rev.\ D {\bf 66}, 015009 (2002)
[arXiv:hep-ph/0201300].

\bibitem{Cheng:2003ju}
H.~C.~Cheng and I.~Low,
JHEP {\bf 0309}, 051 (2003)
[arXiv:hep-ph/0308199].

\bibitem{Cheng:2004yc}
H.~C.~Cheng and I.~Low,
JHEP {\bf 0408}, 061 (2004)
[arXiv:hep-ph/0405243].

\bibitem{Hubisz:2004ft}
J.~Hubisz and P.~Meade,
Phys.\ Rev.\ D {\bf 71}, 035016 (2005)
[arXiv:hep-ph/0411264].

\bibitem{Georgi:2000ks}
H.~Georgi, A.~K.~Grant and G.~Hailu,
Phys.\ Lett.\ B {\bf 506}, 207 (2001)
[arXiv:hep-ph/0012379].

\bibitem{vonGersdorff:2002as}
G.~von Gersdorff, N.~Irges and M.~Quiros,
Nucl.\ Phys.\ B {\bf 635}, 127 (2002)
[arXiv:hep-th/0204223].

\bibitem{Cheng:2002iz}
H.~C.~Cheng, K.~T.~Matchev and M.~Schmaltz,
Phys.\ Rev.\ D {\bf 66}, 036005 (2002)
[arXiv:hep-ph/0204342].

\bibitem{DKM}
A.~Datta, K.~Kong and K.~Matchev, in preparation.

\bibitem{Barr:2004ze}
A.~J.~Barr,
Phys.\ Lett.\ B {\bf 596}, 205 (2004)
[arXiv:hep-ph/0405052].

\bibitem{Servant:2002aq}
G.~Servant and T.~M.~Tait,
Nucl.\ Phys.\ B {\bf 650}, 391 (2003)
[arXiv:hep-ph/0206071].

\bibitem{Arkani-Hamed:2000hv}
N.~Arkani-Hamed, H.~C.~Cheng, B.~A.~Dobrescu and L.~J.~Hall,
Phys.\ Rev.\ D {\bf 62}, 096006 (2000)
[arXiv:hep-ph/0006238].

\bibitem{Appelquist:2002ft}
T.~Appelquist, B.~A.~Dobrescu, E.~Ponton and H.~U.~Yee,
Phys.\ Rev.\ D {\bf 65}, 105019 (2002)
[arXiv:hep-ph/0201131].

\bibitem{Mohapatra:2002ug}
R.~N.~Mohapatra and A.~Perez-Lorenzana,
Phys.\ Rev.\ D {\bf 67}, 075015 (2003)
[arXiv:hep-ph/0212254].

\bibitem{Appelquist:2001mj}
T.~Appelquist, B.~A.~Dobrescu, E.~Ponton and H.~U.~Yee,
Phys.\ Rev.\ Lett.\  {\bf 87}, 181802 (2001)
[arXiv:hep-ph/0107056].

\bibitem{Dobrescu:2001ae}
B.~A.~Dobrescu and E.~Poppitz,
Phys.\ Rev.\ Lett.\  {\bf 87}, 031801 (2001)
[arXiv:hep-ph/0102010].

\bibitem{Agashe:2001ra}
K.~Agashe, N.~G.~Deshpande and G.~H.~Wu,
Phys.\ Lett.\ B {\bf 511}, 85 (2001)
[arXiv:hep-ph/0103235].

\bibitem{Agashe:2001xt}
K.~Agashe, N.~G.~Deshpande and G.~H.~Wu,
Phys.\ Lett.\ B {\bf 514}, 309 (2001)
[arXiv:hep-ph/0105084].

\bibitem{Appelquist:2001jz}
T.~Appelquist and B.~A.~Dobrescu,
Phys.\ Lett.\ B {\bf 516}, 85 (2001)
[arXiv:hep-ph/0106140].

\bibitem{Petriello:2002uu}
F.~J.~Petriello,
JHEP {\bf 0205}, 003 (2002)
[arXiv:hep-ph/0204067].

\bibitem{Appelquist:2002wb}
T.~Appelquist and H.~U.~Yee,
Phys.\ Rev.\ D {\bf 67}, 055002 (2003)
[arXiv:hep-ph/0211023].

\bibitem{Chakraverty:2002qk}
D.~Chakraverty, K.~Huitu and A.~Kundu,
Phys.\ Lett.\ B {\bf 558}, 173 (2003)
[arXiv:hep-ph/0212047].

\bibitem{Buras:2002ej}
A.~J.~Buras, M.~Spranger and A.~Weiler,
Nucl.\ Phys.\ B {\bf 660}, 225 (2003)
[arXiv:hep-ph/0212143].

\bibitem{Buras:2003mk}
A.~J.~Buras, A.~Poschenrieder, M.~Spranger and A.~Weiler,
Nucl.\ Phys.\ B {\bf 678}, 455 (2004)
[arXiv:hep-ph/0306158].

\bibitem{Dienes:1998vg}
K.~R.~Dienes, E.~Dudas and T.~Gherghetta,
Nucl.\ Phys.\ B {\bf 537}, 47 (1999)
[arXiv:hep-ph/9806292].

\bibitem{Cheng:2002ej}
H.~C.~Cheng, J.~L.~Feng and K.~T.~Matchev,
Phys.\ Rev.\ Lett.\  {\bf 89}, 211301 (2002)
[arXiv:hep-ph/0207125].

\bibitem{Hooper:2002gs}
D.~Hooper and G.~D.~Kribs,
Phys.\ Rev.\ D {\bf 67}, 055003 (2003)
[arXiv:hep-ph/0208261].

\bibitem{Servant:2002hb}
G.~Servant and T.~M.~Tait,
New J.\ Phys.\  {\bf 4}, 99 (2002)
[arXiv:hep-ph/0209262].

\bibitem{Majumdar:2002mw}
D.~Majumdar,
Phys.\ Rev.\ D {\bf 67}, 095010 (2003)
[arXiv:hep-ph/0209277].

\bibitem{Bertone:2002ms}
G.~Bertone, G.~Servant and G.~Sigl,
Phys.\ Rev.\ D {\bf 68}, 044008 (2003)
[arXiv:hep-ph/0211342].

\bibitem{Hooper:2004xn}
D.~Hooper and G.~D.~Kribs,
Phys.\ Rev.\ D {\bf 70}, 115004 (2004)
[arXiv:hep-ph/0406026].

\bibitem{Bergstrom:2004cy}
L.~Bergstrom, T.~Bringmann, M.~Eriksson and M.~Gustafsson,
Phys.\ Rev.\ Lett.\  {\bf 94}, 131301 (2005)
[arXiv:astro-ph/0410359].

\bibitem{Macesanu:2002ew}
C.~Macesanu, C.~D.~McMullen and S.~Nandi,
Phys.\ Lett.\ B {\bf 546}, 253 (2002)
[arXiv:hep-ph/0207269].

\bibitem{Cheng:2002rn}
H.~C.~Cheng,
Int.\ J.\ Mod.\ Phys.\ A {\bf 18}, 2779 (2003)
[arXiv:hep-ph/0206035].

\bibitem{Pukhov:1999gg}
A.~Pukhov {\it et al.},
arXiv:hep-ph/9908288.

\bibitem{Sjostrand:2001yu}
T.~Sjostrand, L.~Lonnblad and S.~Mrenna,
arXiv:hep-ph/0108264.

\bibitem{Pohl:2002vk}
M.~Pohl and H.~J.~Schreiber,
arXiv:hep-ex/0206009.

\bibitem{peskin}
M.~Peskin,
``Consistent Yokoya-Chen approximation to beamstrahlung,''
SLAC-TN-04-032.

\bibitem{Battaglia:2001dg}
M.~Battaglia, S.~Jadach and D.~Bardin,
in {\it Proc. of the APS/DPF/DPB Summer Study on the Future of Particle Physics 
(Snowmass 2001)}, ed. N.~Graf, eConf {\bf C010630} (2001) E3015.

\bibitem{blair}
G.~A.~Blair,
in {\it Proc. of the APS/DPF/DPB Summer Study on the Future of Particle Physics (Snowmass 2001)},
ed. N.~Graf, eConf {\bf C010630}, E3019 (2001).

\bibitem{Battaglia:2002eg}
M.~Battaglia and M.~Gruwe,
arXiv:hep-ph/0212140.

\bibitem{Battaglia:2003wh}
M.~Battaglia,
Nucl.\ Instrum.\ Meth.\ A {\bf 522} (2004) 19
[arXiv:hep-ex/0311041].

\bibitem{Bhattacharyya:2005vm}
G.~Bhattacharyya, P.~Dey, A.~Kundu and A.~Raychaudhuri,
arXiv:hep-ph/0502031.

\end{thebibliography}
\end{document}